\documentstyle[12pt,epsf]{article}
\def\beq{\begin{equation}}
\def\eeq{\end{equation}}
\def\bea{\begin{eqnarray}}
\def\eea{\end{eqnarray}}
\def\R{\rangle}
\def\L{\langle}
\def\lt{\left}
\def\rt{\right}
\newcommand{\sectiono}[1]{\section{#1}\setcounter{equation}{0}}
\newcommand{\subsectiono}[1]{\subsection{#1}}

\begin{document}
{}~
\hfill\vbox{\hbox{hep-th/0110136}\hbox{MRI-P-011002} \hbox{PSU/TH-247} 
\hbox{NSF-ITP-01-159}
}\break

\vskip 1.4cm

\centerline{\large \bf Oscillator Representation of the BCFT Construction}
\centerline{\large \bf of D-branes in}
\centerline{\large \bf Vacuum String Field Theory}

\vspace*{5.0ex}
\centerline{\large \rm Partha Mukhopadhyay}

\vspace*{6.5ex}

\centerline{\large \it Harish-Chandra Research
Institute\footnote{Formerly Mehta Research Institute of Mathematics
and Mathematical Physics}}

\centerline{\large \it  Chhatnag Road, Jhusi,
Allahabad 211019, INDIA}

\vspace*{2ex}

\centerline{\large and}

\vspace*{2ex}

\centerline{\large \it Department of Physics}

\centerline{\large \it Penn State University}

\centerline{\large \it University Park, PA 16802, USA}
        
\centerline{e-mail: partha@mri.ernet.in, partha@phys.psu.edu}

\vspace*{8.5ex}

\centerline{\bf Abstract}
\bigskip
Starting from the boundary CFT definition for the D-branes in vacuum string 
field theory (VSFT) given in hep-th/0105168, we derive the oscillator 
expression for 
the D24-brane solution in the VSFT on D25-brane. We show that the state takes 
the form of a squeezed state, similar to the one found directly in terms of 
the oscillators and reported in hep-th/0102112. Both the solutions are actually
 one parameter families of solutions. We also find numerical evidence 
that at least for moderately large values of the parameter $(b)$ in the 
oscillator construction the two families of solutions are same under a 
suitable redefinition of the parameter. Finally we generalize the method to 
computing the oscillator expression for a D-brane solution with constant gauge 
field strength turned on along the world volume. 

\vfill \eject

\baselineskip=17.3pt

\tableofcontents

%
%
\sectiono{Introduction} \label{intro}

Since it was realized \cite{universality} that a string field 
theory \cite{W,BER} can provide a 
useful setup for the study of Sen's conjectures, much work has taken 
place to prove the conjectures using Witten's cubic string field theory 
(CSFT)\cite{W} in the context of bosonic open string theory. A large body of 
numerical work
\cite{KoSa,SeZw,MoTa,HvKr,MeJMT,MoSeZw,RaZw,SeZw2,Ta,MeRod,Mo,HaSh,Zw,Sch,MuSe,HaTe,EllTa,FeHMo,Oh,EllFHM} has verified the conjectures to high accuracy and 
helped gathering experience on CSFT itself. 

Very recently, mainly to understand the part of the conjectures involving the 
nonexistence of the physical open string excitations around the tachyon vacuum,
Rastelli, Sen and Zwiebach (RSZ) started the programme of attacking the problem
analytically\cite{RSZ1,RSZ2,RSZ3,RSZ4}. In this work\cite{RSZ1} they 
conjectured a simple 
form for the CSFT expanded
 around the tachyon vacuum, although it could not be derived from first 
principles because of the lack of 
knowledge of the vacuum solution in an analytic form. This new string field 
theory which has been called vacuum string field theory (VSFT)\footnote{See 
\cite{GT1, GT2, kawano, david, furuuchi, hata-kawano, kishimoto}
for related investigations. }in the 
literature, possesses exactly the same structure as CSFT with the only 
exception
 that the BRST operator $Q_B$ in the original CSFT is now replaced by an 
unknown operator $\cal{Q}$ depending only on the ghost part of the 
theory\footnote{The operator $Q_B$ being unknown reflects the fact that the 
complete 
vacuum solution in CSFT is lacking.}. Various important algebraic equations in 
CSFT take the 
same form in VSFT with $Q_B$ replaced by $\cal{Q}$ and hence VSFT has the same
gauge invariance and equation of motion as that of CSFT with $Q_B$ replaced by
$\cal{Q}$.

This work was followed by another\cite{RSZ2} by the same authors where the 
matter parts of
various classical solutions of VSFT on D25 brane were constructed using the
oscillator language with an 
ansatz of factorization between the ghost and matter parts. Since these are 
lump solutions of various codimensions and they produce the expected ratio of 
tensions of D-branes to a reasonable numerical accuracy, the solutions were 
interpreted as D-branes of 
various world volume dimensions. Each of these solutions (except for the one 
with zero codimension representing the D25 brane itself) is labelled by a 
continuous parameter $b$. This has been interpreted to be a gauge parameter,
which means that all the solutions in each one parameter family are actually 
gauge transforms of each other. Existence of these solutions gave the RSZ 
conjecture a strong foundation. 

After this, various works\cite{RSZ3,GT1,GT2} have been done towards finding 
the multi-D-brane 
solutions in VSFT. But in the present paper we will be interested in the 
recent work\cite{RSZ4} by RSZ where a different construction of the D-brane 
solutions in any VSFT has been given through a boundary conformal field theory
(BCFT) prescription. BCFT techniques have been used to show 
that the solutions really satisfy the VSFT equation of motion. In this approach
the matter part of any D-brane solution in the VSFT on that D-brane itself is
the matter part of the sliver (which is a surface state)
\cite{RaZw,RSZ2,RSZ3,RSZ4} corresponding to the
relevant BCFT. The matter part of any other D-brane solution has been given as
the matter part of a surface-like state where certain boundary condition 
changing twist operators appear in the correlation functions defining the 
state. The points of insertions of these operators involve a continuous
parameter $\epsilon$. It has been shown in ref.\cite{RSZ4} that the states 
satisfy the desired 
equation of motion and produce the correct ratio of tensions for an arbitrary 
positive value of $\epsilon$. Hence it has been interpreted as a gauge 
parameter like $b$ in the oscillator solutions.  

In this work we will be interested in finding the connection between the two 
one parameter families of solutions corresponding to a D24-brane in the VSFT
on D25-brane. For this we first find the oscillator expression for the 
solution described through BCFT construction\cite{RSZ4}. The state takes the 
form of a squeezed state, - same structure as the one found in oscillator 
description\cite{RSZ2}. Moreover we find numerical evidence that the two 
families of solutions are actually same with $\epsilon $
 being a specific function of $b$. Our computational method relies heavily 
on the application of Wick's theorem which holds as long as the boundary 
condition on the fields is linear. We also indicate that the same method 
can be applied to find the solution for a D-brane with constant gauge field
strength turned on along the world volume.

The organization of the paper is as follows: In sec.\ref{review} we review the 
oscillator and BCFT constructions of the D-brane solutions in the VSFT on 
D25-brane. For the sake of simplicity we will restrict ourselves to the 
solution of D24-brane placed at the origin with the 25th direction being 
transverse to the brane. In sec.\ref{D24-oscillator} we will start from the 
BCFT construction of the solution describing the same D24-brane and derive 
its oscillator expression. In sec.\ref{numerics} we will present the numerical 
result showing that the $\epsilon$ and $b$-families of solutions are same with 
$\epsilon$ being a specific function of $b$ for moderately large values of 
$b$, namely $10-25$. In sec.\ref{generalization} we will discuss the more 
generic 
situation where a constant gauge field strength is turned on along the world 
volume. We discuss some numerical issues especially for small values of $b$
in sec.\ref{discussion}. Here we show some numerical features which imply that
the solutions might be same even for smaller values of $b$. Some of the 
computational details are given in the  appendices.   

%
%
\sectiono{Review of the oscillator and BCFT construction of the D-brane 
solutions}\label{review}   

Let us start with the classical equation of motion in VSFT. 
\bea
{\cal{Q}} |\Psi\R + |\Psi * \Psi \R = 0. \label{eom}
\eea
where the string field $\Psi$ is a ghost number 1 state in the Hilbert space
(denoted ${\cal{H}}_{BCFT}$) 
of the combined matter-ghost boundary conformal field theory corresponding to 
the D-brane on which the string field theory is considered. The full BCFT can 
be written in the following way:
\bea
BCFT = \lt(BCFT\rt)_m \oplus \lt(BCFT\rt)_g, \label{BCFT}
\eea
where the subscripts $m$ and $g$ refer to matter and ghost respectively. 
$\lt(BCFT\rt)_g$ is the usual $bc$ CFT on the upper half plane with central 
charge $-26$. This part is universal in the sense that 
boundary conditions of the ghost fields is same for all the D-branes. 
$\lt(BCFT\rt)_m$ is the direct sum of 26 CFT's on the upper half plane, each 
consisting of a single scalar field with boundary condition depending on the 
D-brane considered.  $\cal{Q}$ which replaces the BRST charge $Q_B$ in the 
original CSFT, depends purely on the ghost operators. The star product $A*B$ 
has its usual meaning. 

\noindent {\bf The factorization ansatz:}
The factorization ansatz\cite{RSZ2} for the D-brane solutions to the above equation reads,
\bea
|\Psi\R = |\Psi_g\R \otimes |\Psi_m\R, \label{factorization}
\eea
where $|\Psi_g\R$ depends only on the ghosts and is common to
all the D-brane solutions. $|\Psi_m\R$ is the matter part which varies for 
different D-branes. 

There is an ambiguity in fixing the 
overall normalizations of $|\Psi_g\R$ and $|\Psi_m\R$. This can be fixed by 
demanding that the factorized equations of motion take the following 
forms\cite{RSZ2}:
\bea
{\cal{Q}} |\Psi_g\R &=& -|\Psi_g *^g \Psi_g \R, \\ \label{factorized-eom1}
|\Psi_m\R &=& |\Psi_m *^m \Psi_m \R, \label{factorized-eom2}  
\eea
where $*^g$ and $*^m$ denote respectively the ghost and matter parts of the 
star product. Although the normalizations of the ghost and matter parts can 
diverge or become zero because of the expected appearance of the conformal 
anomaly while performing computations in either the ghost or the matter 
conformal field theory, the full solution is expected to be well-behaved.    

Here we will be concerned only with eq.(\ref{factorized-eom2}), the solutions 
to which have been found in ref.\cite{RSZ2,RSZ3,RSZ4,GT1}. Below we will 
consider the VSFT on D25 brane and 
discuss the D24-brane solution in oscillator\cite{RSZ2} and BCFT\cite{RSZ4}
language.

%
%
\subsection{Oscillator description of the D-brane solutions}
\label{oscillator-construction} 

We will start with the space-time independent solution\cite{KoSa,RSZ2}
 corresponding to the 
D25-brane. This is given by,
\bea
|\Psi_m\R = {\cal{N}}^{26}\exp\lt(-\frac{1}{2} \eta_{\mu \nu} \sum_{m,n \geq 1}
S_{mn} a^{\mu \dagger}_m a^{\nu \dagger}_n \rt) |0_{26}\R. \label{D25}
\eea
Here we have adopted the same notation of ref.\cite{RSZ2} except for the state 
$|0_{26}\R$. This is the SL(2,R) invariant 
vacuum which corresponds to the zero value for the 26-dimensional momentum. 
In this notation a state $|0_n\R$ is normalized as follows: $\L0_n|0_n\R= 
{V^{(n)} \over (2\pi)^n}$, $V^{(n)}$ being the volume of the $n$-dimensional 
Minkowski space.
The normalization constant $\cal{N}$ and the infinite dimensional matrix
 $S$ are given by the following relations\cite{RSZ2}:
\bea
&\cal{N}& = \det\lt(1-X\rt)^{1/2} \det\lt(1+T\rt)^{1/2},~~~ S= CT, \cr
\cr
&T& =\frac{1}{2X}\lt(1+X - \sqrt{\lt(1+3X\rt)\lt(1-X\rt)}\rt), \cr
\cr
&X& =CV^{11},~~~C_{mn}= (-1)^m \delta_{mn}. \label{NSTXC}
\eea
Here $V^{11}$ is one of the infinite dimensional matrices $V^{rs}$ with 
$r,s=1,2,3$\cite{cremmer,Sam,ohta,GJ1,GJ2,RSZ2} that appear in the three-string
 vertex in the oscillator formalism.
For definition and important algebraic properties of these matrices the reader 
is referred to \cite{RSZ2}. 

Let us now consider the D24-brane solution placed at the origin with $x^{25}$
as the transverse coordinate. The state looks exactly like the state in 
eq.(\ref{D25}) for the tangential directions as $\lt(BCFT\rt)_m$ is the 
decoupled theory of 26 scalar fields. The $X\equiv X^{25}$ part will now have 
momentum dependence. The solution takes the following form:
\bea
|\Psi_m^{\prime}(b)\R = {\cal{N}}^{25} \exp\lt(-\frac{1}{2} 
\eta_{\bar\mu  \bar\nu} \sum_{m,n \geq 1}
S_{mn} a^{\bar \mu \dagger}_m a^{\bar \nu \dagger}_n \rt) |0_{25}\R \otimes
|\Psi_X^{\prime}(b)\R, \label{Psi-prime}
\eea
where $\bar \mu , \bar \nu = 0,1\cdots 24$ and, 
\bea
|\Psi_X^{\prime}(b)\R ={\cal{N}^{\prime}} \exp\lt(-\frac{1}{2} 
\sum_{m,n \geq 0}S_{mn}^{\prime} a^{\dagger}_m a^{\dagger}_n \rt) |\Omega_b\R,
\label{Psip_X1} 
\eea
where the zero mode oscillator is defined through:
\bea
a_0 = 
{\sqrt{b} \over 2} \hat{p}~-~{i \over \sqrt{b}} \hat{x}, ~~~~~
a^{\dagger}_0 = 
{\sqrt{b} \over 2} \hat{p}~+~ {i \over \sqrt{b}} \hat{x}.
\eea
Here we have suppressed the superscript ``$25$'' for the 25-th direction. 
$b$ is an arbitrary real number. The constant 
$\cal{N}^{\prime}$ and the infinite dimensional matrix $S^{\prime}$
 are given as follows\cite{RSZ2}:
\\
\\
\\
\\
\bea
&&{\cal{N}}^{\prime} = {\sqrt 3 \over (2\pi b^3)^{1/4}}
\lt(V^{rr}_{00}+{b\over 2}\rt) \{\det(1 - X')^{1/2}\det(1+T')^{1/2} \},  \cr
\cr
&&S^{\prime} = C^{\prime}T^{\prime},~~~ 
T^{\prime} = {1\over 2X^{\prime}}\lt(1+X^{\prime} - \sqrt{\lt(1+3X^{\prime}\rt)
\lt(1-X^{\prime}\rt)}\rt),  \cr
\cr
&&X^{\prime} = C^{\prime}V^{\prime 11}(b),~~~C_{mn}^{\prime}= (-1)^m 
\delta_{mn}, ~~~ V^{rr}_{00}= \ln(27/16).	
\label{NSTXCp}
\eea
$V^{\prime rs}(b)$ are the $b$ dependent 
matrices introduced in ref.\cite{RSZ2} which have the similar algebraic 
properties as 
$V^{rs}$ matrices. The relation between the $\hat{p}$ eigenstate and the 
normalized zero mode harmonic oscillator ground state $|\Omega_b\R$ is given by:
\bea
|p\R = \lt(2\pi/b\rt)^{-1/4}~\exp\lt[-\frac{b}{4}~p^2 ~+~ \sqrt{b}~p~
a_0^{\dagger} ~-~ \frac{1}{2}\lt(a_0^{\dagger}\rt)^2\rt] |\Omega_b\R.
\eea
Using this relation one can express eq.(\ref{Psip_X1}) in terms of the 
momentum eigenstates. It takes the following form:
\bea
|\Psi^{\prime}_X(b)\R = \int dp~ h(p,b) ~ \exp \lt(-~{1\over 2}~ 
a^{\dagger}.Q(b).a^{\dagger}~+~p~ L(b).a^{\dagger} \rt)~ |p\R, 
\label{Psip_X2}
\eea
where,
\bea
Q_{mn}(b) &=& S^{\prime}_{mn} ~+~ {S^{\prime}_{0m}S^{\prime}_{0n} \over
(1-S^{\prime}_{00})}, ~~~~~~~~~~~~\forall m,n \geq 1, \cr
\cr
L_n(b) &=& -{\sqrt{b}S^{\prime}_{0n} \over 1-S^{\prime}_{00} }, 
~~~~~~~~~~~~~~~~~~~~~~~\forall n\geq 1, \cr
\cr 
h(p,b) &=& \lt(2\pi /b\rt)^{-1/4}~(1-S^{\prime}_{00})^{-1/2}
{\cal{N}}^{\prime}\exp \lt(-{B\over 2}p^2 \rt), \cr
\cr
B &=& b \lt( {1\over 2} ~+~ {S^{\prime}_{00} \over 1 - S^{\prime}_{00}} \rt).
\label{QLhB} 
\eea
%
\subsection{BCFT construction of the D-brane solutions}
\label{BCFT-construction}
%
In ref.\cite{RSZ4} the D-brane solutions of a VSFT have been constructed using 
the BCFT
description. In this approach every D-brane solution is described by the
sliver $|\Xi_{BCFT}\R$ of the corresponding BCFT. An open string field theory
is always defined on a particular D-brane. Let us denote the BCFT and the 
state space corresponding to this 
reference D-brane by $BCFT$ and ${\cal{H}}_{BCFT}$ respectively. Now if
$BCFT^{\prime}$ and ${\cal{H}}_{BCFT^{\prime}}$ denote respectively the BCFT 
and 
the state space 
corresponding to the D-brane for which the solution is sought, then the matter 
part of the solution is given by the matter part of $|\Xi_{BCFT^{\prime}}\R$ 
expressed 
in ${\cal{H}}_{BCFT}$ \footnote{Note that $|\Xi_{BCFT^{\prime}}\R$ is 
originally a 
state in ${\cal{H}}_{BCFT^{\prime}}$.}. The prescription for expressing 
$|\Xi_{BCFT^{\prime}}\R$
in ${\cal{H}}_{BCFT}$ is given by giving the BPZ inner product of 
$|\Xi_{BCFT^{\prime}}\R$ with an arbitrary state in ${\cal{H}}_{BCFT}$. This is
given by the following correlator on the UHP \cite{RSZ4}:
\bea
\L\Xi_{BCFT^{\prime}}|\phi\R = (2\epsilon)^{2h} \left \L 
f\circ \phi(0) ~\sigma^+ \lt( \frac{\pi}{4}+\epsilon \rt) 
\sigma^- \lt( -\frac{\pi}{4}-\epsilon \rt) \right \R_{BCFT}, ~~~~
|\phi\R \in {\cal{H}}_{BCFT}, \label{sliver-defined}
\eea
where $\sigma^+(t)~(t \in \bf{R})$ is the vertex operator of the ground state
of string with its left end (if viewed from inside the UHP) on the 
$BCFT^{\prime}$-brane and right end on the $BCFT$-brane. Similarly 
$\sigma^-(t)$ is the ground state vertex operator for the string connecting the
 two branes in the opposite orientation. These operators\footnote{For a 
systematic study of the correlation functions in presence of the $\sigma $ 
operators in case of only Neumann and Dirichlet boundary conditions and for
more references on this subject see 
ref.\cite{dpdq}.} are dimension $h$ 
primaries which change the boundary condition respectively from $BCFT$ to 
$BCFT^{\prime}$ and $BCFT^{\prime}$ to $BCFT$ on the real line. Therefore the
above notation means that the correlator has to be computed with $BCFT$ 
boundary 
condition for $|t| \leq (\frac{\pi}{4}+\epsilon)$ and $BCFT^{\prime}$ boundary 
condition on the rest of the real line. $\epsilon$ is an arbitrary real 
positive 
parameter. $\phi(t)$ is the vertex operator for the state $|\phi\R$ and 
$f\circ \phi(t)$ denotes the transformation of $\phi(t)$ under the following
conformal map:
\bea
f(z) = \tan^{-1}(z).
\eea 
In case $BCFT^{\prime}$ and $BCFT$ are same, $\sigma^{\pm}$ become the 
identity
operator with $h=0$ and the above definition reduces to the usual definition of
the sliver \cite{RSZ4}. 

It has been demonstrated in ref.\cite{RSZ4} that $|\Xi_{BCFT^{\prime}}\R$ 
$*$-multiplies to itself for any $\epsilon$. From the definition 
(\ref{sliver-defined}) it is clear that 
$|\Xi_{BCFT^{\prime}}\R$ factorizes into ghost and matter parts since 
$\sigma^{\pm}$ do not involve any ghost part.
But again the ambiguity
of normalization is obvious. The normalization can be fixed by demanding that
the matter part $*$-multiplies to itself:
\bea
|\Xi_{BCFT^{\prime}_m}\R *^m |\Xi_{BCFT^{\prime}_m}\R = 
|\Xi_{BCFT^{\prime}_m} \R.
\eea
$|\Xi_{BCFT^{\prime}_m}\R$ defined in this way is conjectured to be the matter 
part of the $BCFT^{\prime}$-brane solution.
%
%
\sectiono{Oscillator expression for D24-brane solution in BCFT construction}
\label{D24-oscillator}
Here we will use the prescription (\ref{sliver-defined}) to find the oscillator
expression for the D24-brane solution in the VSFT formulated on D25-brane. The
D24-brane will be placed at the origin with $x^{25}$ as the
transverse direction. 
Therefore in our analysis $BCFT$ will correspond to the D25-brane and 
$BCFT^{\prime}$ to the D24-brane. For 
notational simplicity we will denote $|\Xi_{BCFT}\R$ by $|\Xi\R$ and  
$|\Xi_{BCFT^{\prime}}\R$ by $|\Xi^{\prime}\R$.

The first step in our computation will be to find the BCFT construction for 
the matter part of $BCFT^{\prime}$ sliver i.e. $|\Xi^{\prime}_m\R$. It 
is given by a prescription similar to that in eq.(\ref{sliver-defined})
 with $\phi \in {\cal{H}}_{BCFT_m}$. The only 
difference is that, now there is an unknown overall constant which is so 
adjusted that $|\Xi^{\prime}_m\R$ (and hence also $|\Xi^{\prime}_g\R$) 
$*$-multiplies \footnote{Note that the star
product of $|\Xi^{\prime}_m\R$ can not be computed using the nice geometrical 
arguments used in ref.\cite{RSZ4} because of the conformal anomaly coming from 
the nonzero central charge of $BCFT_m$.} to itself. Therefore,
\bea
\L\Xi^{\prime}_m|\phi\R = \widehat{\cal{N}}^{26} (2\epsilon)^{2h} \left\L 
f\circ \phi(0)~ \sigma^+ \lt(\frac{\pi}{4}+\epsilon \rt) 
 \sigma^- \lt(-\frac{\pi}{4}-\epsilon \rt) \right \R_{BCFT_m}, ~~~~~ 
|\phi\R \in {\cal{H}}_{BCFT_m}. \label{sliverp-m-defined}
\eea
$\widehat {\cal N}$ can be found a follows. 
Since $|\Xi^{\prime}_g\R \otimes |\Xi^{\prime}_m \R $ is given by 
eq.(\ref{sliver-defined}), $|\Xi^{\prime}_g\R $ will have a normalization 
constant $\widehat{\cal{N}}^{-26}$. On the other hand $|\Xi^{\prime}_g\R $ and
$|\Xi_g\R $ are identical, and hence $|\Xi_m\R $ must have the same 
normalization constant $\widehat{\cal{N}}^{26}$.
\bea
\L\Xi_m|\phi\R = \widehat{\cal{N}}^{26} \L f\circ \phi(0) \R_{BCFT_m}, 
~~~~~~~~ |\phi\R \in {\cal{H}}_{BCFT_m}. \label{sliver-m-defined}
\eea 
It can be proved (see end of this section)
that $|\Xi_m\R$ with the above
definition has the following oscillator expression:
\bea
|\Xi_m\R = \widehat{\cal{N}}^{26}\exp\lt(-\frac{1}{2} \eta_{\mu \nu} 
\sum_{m,n \geq 1} \widehat S_{mn} a^{\mu \dagger}_m a^{\nu \dagger}_n \rt) 
|0_{26}\R, \label{sliver-oscillator}
\eea
where $\widehat S_{mn}$'s are given as follows \cite{RSZ2},
\bea
\widehat S_{mn} &=& -{1\over \sqrt{mn}} \oint {dz \over 2\pi i} 
\oint {dw \over 2\pi i}~ (\tan z)^{-m}~ (\tan w)^{-n}~ (z-w)^{-2}.
\label{S_mn}
\eea
It was argued with numerical results in 
ref.\cite{RSZ2} that 
(see eqs.(\ref{D25}), (\ref{NSTXC}))
\bea
{\widehat{\cal{N}}} = {\cal{N}}, ~~&&~~ \widehat S_{mn}=S_{mn}.
\eea
Thus,
\bea
\L\Xi^{\prime}_m|\phi\R = {\cal{N}}^{26} (2\epsilon)^{2h} \left\L 
f\circ \phi(0) ~\sigma^+ \lt(\frac{\pi}{4}+\epsilon \rt) 
 \sigma^- \lt(-\frac{\pi}{4}-\epsilon \rt) \right \R_{BCFT_m}, ~~~~~
|\phi\R \in {\cal{H}}_{BCFT_m}. \label{sliverp-m-defined2}
\eea

Now the difference between $BCFT$ and $BCFT^{\prime}$ comes only from the 
world-sheet field $X = X^{25}$ as this is the only one which has N-D 
boundary condition. All the other matter fields are N-N. Therefore only $X$ 
can see the presence of $\sigma^{\pm}$ and it acts as the identity operator for
 all 
the other
matter fields. Hence $|\Xi^{\prime}_m\R$ looks exactly like $|\Xi_m\R$ for the
excitations along $X^0,\cdots, X^{24}$. Therefore,
\bea
|\Xi^{\prime}_m(\epsilon)\R = {\cal{N}}^{25}\exp\lt(-\frac{1}{2} \eta_{\bar
\mu \bar \nu} \sum_{m,n \geq 1} S_{mn} a^{\bar \mu \dagger}_m a^{\bar \nu 
\dagger}_n \rt) |0_{25}\R \otimes |\Xi^{\prime}_X(\epsilon)\R,
\label{Xip_m}
\eea
where $|\Xi^{\prime}_X(\epsilon)\R$ is defined through the following relation:
\bea
\L\Xi^{\prime}_X(\epsilon)|\phi\R &=& {\cal{N}} ~ (2\epsilon)^{2h} \L f\circ 
\phi(0) ~ \sigma^+(t_0) ~ \sigma^-(-t_0) \R_{BCFT_X}, ~~~~~~ 
|\phi\R \in {\cal{H}}_{BCFT_X}, \cr
t_0 &=& \frac{\pi}{4} + \epsilon. \label{Xip_X1}
\eea

Hence at this stage our job is to find the oscillator expression for 
$|\Xi^{\prime}_X(\epsilon)\R$ and compare it with eq.(\ref{Psip_X2}). To this 
end we find it useful to define\footnote{We thank Justin David for his 
suggestion on this issue.} the following 
generating function,\footnote{This point onwards we will suppress the 
subscript ``$BCFT_X$''in the correlation function for notational simplicity.}
\bea
G(j, p, t_0) \equiv \L f \circ \exp(j.a^{\dagger}) ~ e^{ipX}(0) ~ 
\sigma^+ (t_0) ~ \sigma^- (-t_0)\R, \label{G1}
\eea
where 
$j.a^{\dagger}=\sum_{n\geq 1}j_n a^{\dagger}_n$ and $j_n$'s are real 
numbers. $f \circ \exp(j.a^{\dagger})$ is defined as follows. The residue
expression for the creation operators in the $\alpha^{\prime}=1$ unit is given 
by:
\bea
a^{\dagger}_n = \frac{1}{\sqrt{2n}}\ointop {dw \over 2\pi}~w^{-n}~\partial X(w)
. \label{adagger}
\eea
Now applying the technique of ref.\cite{LPP1,LPP2} we get,
\bea
f \circ a^{\dagger}_n = \frac{1}{\sqrt{2n}}\ointop {dz \over 2\pi}~\lt(
f^{-1}(z)\rt)^{-n} ~\partial X(z).  \label{fdotadagger}
\eea
Then we define $f \circ \exp(j.a^{\dagger})$ through the power series 
expansion of the exponential.
\bea
f \circ \exp(j.a^{\dagger}) = \sum_{N \geq 0} {1\over N!} \lt[
j.\lt(f \circ a^{\dagger}\rt)\rt]^{N}.
\eea
The use of $G(j, p, t_0)$ lies in the fact that it directly gives the 
oscillator expression for $|\Xi^{\prime}_X\R$ through the following 
prescription (see appendix \ref{DXip_X2} for a proof):
\bea
|\Xi^{\prime}_X(\epsilon)\R = {\cal{N}}~(2\epsilon)^{2h}~\int dp ~G(j_n 
~\rightarrow (-1)^{n+1}~a^{\dagger}_n, -p, t_0)~|p\R, \label{Xip_X2}
\eea
where $G(j_n \rightarrow (-1)^{n+1}a^{\dagger}_n, -p, t_0)$ in the above 
expression means that $j_n$ is replaced by $(-1)^{n+1}~a^{\dagger}_n$ in the 
expression of $G(j, -p, t_0)$ after its computation is over. Therefore our next
job will be to compute the generating function $G(j, p, t_0)$. 

%
%
\noindent {\bf Computation of $G(j, p, t_0)$:}
Our approach of computing $G(j, p, t_0)$ will be to find a first order 
differential equation of $G(j, p, t_0)$ with respect to $j_n$ and then 
integrate it. Let us start by expanding $\exp(j.a^{\dagger})$ in eq.(\ref{G1})
\bea
&G(j, p, t_0)& =\sum_{N=0}^{\infty}~ \frac{1}{N!}~ \L 
(j.f\circ a^{\dagger})^N~ e^{ipX}(0)~\sigma^+ (t_0) ~ \sigma^- (-t_0)\R \cr
\cr
&=&\sum_{N=0}^{\infty}~ \frac{1}{N!}\sum_{\{n_i\}} \prod_{i=1}^N \lt( j_{n_i}
\L f\circ a^{\dagger}_{n_i}~ e^{ipX}(0)~ \sigma^+ (t_0) ~ \sigma^- (-t_0)\R 
\rt)\cr
\cr
&=&\sum_{N=0}^{\infty} {1\over N!} \sum_{\{n_i\}} \prod_{i=1}^N
\lt( {j_{n_i} \over \sqrt{2n_i}}\rt) \oint ~\prod_{i=1}^N \lt[ 
\frac{dz_i}{2\pi}~ (\tan z_i)^{-n_i} \rt] \cr
&&\L \prod_{i=1}^N \partial X(z_i)~ 
e^{ipX}(0)~ \sigma^+ (t_0) ~ \sigma^- (-t_0)\R, \label{G2} 
\eea
where in the last step we have used eq.(\ref{fdotadagger}) with 
$f^{-1}(z)=\tan z$. Similarly the first order derivative with respect to 
$j_m$ takes the following form:
\bea
\partial_{j_m}G(j, p, t_0) &=& \sum_{N=0}^{\infty} {1\over N!} \sum_{\{n_i\}} 
\prod_{i=1}^N \lt( {j_{n_i} \over \sqrt{2n_i}}\rt) \frac{1}{\sqrt{2m}} 
\oint ~\prod_{i=1}^N \lt[ \frac{dz_i}{2\pi}~ (\tan z_i)^{-n_i} \rt] \cr
&&\oint {dz\over 2\pi}~ (\tan z)^{-m} \L \partial X(z) \prod_{i=1}^N 
\partial X(z_i)~ e^{ipX}(0) \R_{\sigma(t_0)},   \label{dG1}
\eea
where we have used the following notation: $\L \cdots \R_{\sigma(t_0)}
\equiv \L \cdots \sigma^+ (t_0) ~ \sigma^- (-t_0)\R $. Now our aim will be to 
extract the $z$ dependence of the above correlation function so that the $z$
integral can finally be taken out of all the summations. This could be done by
 computing the full correlation function. But we find it easier to do the 
computation in a different way. Our aim is to express the right hand side 
in such a way that it has $G(j, p, t_0)$ as a factor. Once it is done we can 
do the integration quite easily. This will be achieved by explicitly writing 
the
result of contracting only $\partial X(z)$ with the other fields and leaving 
the other parts uncontracted\footnote{This can only be done in cases where 
Wick's theorem holds. Although the presence of $\sigma $ operators means that 
the boundary theory is interacting we will explicitly show 
in appendix \ref{sigma} that Wick'e theorem actually holds, but with an 
effective normalized two point function.}. Doing this we get,
\bea
\L \partial X(z) \prod_{i=1}^N \partial X(z_i)~ e^{ipX}(0) \R_{\sigma(t_0)}
&=& \sum_{k=1}^N  T_2(z,z_k,t_0)~ \L \prod_{i\neq k}^N \partial 
X(z_i)~e^{ipX}(0) \R_{\sigma(t_0)} \cr
\cr 
&& + ~ip~T_1(z,t_0)~ \L \prod_{i=1}^N \partial 
X(z_i)~e^{ipX}(0) \R_{\sigma(t_0)}. \label{dX-dX-e}
\eea
Here,
\bea
T_2(z,z_k,t_0) \equiv  \partial_z \partial_{z_k} {\cal G}^N(z, z_k, t_0), &&
T_1(z,t_0) \equiv  \partial_z {\cal G}^N(z,0, t_0)
\label{T2T1}
\eea
are the results of contracting $\partial X(z)$ with $\partial X(z_k)$ 
and  $X(0)$ respectively. ${\cal G}^N$ is an effective normalized two point 
function defined through\footnote{The details of calculations of the 
correlators
in presence of the $\sigma $ operators \cite{dpdq} will be presented in 
appendix \ref{sigma}. },
\bea
{\cal G}^N(z, w, t_0) &\equiv& {\L X(z)~X(w)~\sigma^+(t_0)~\sigma^-(-t_0)\R 
\over \L\sigma^+(t_0)~\sigma^-(-t_0)\R}.
\eea
Using eq.(\ref{dX-dX-e}) in eq.(\ref{dG1}) one gets,
\bea
&&\partial_{j_m}G(j, p, t_0) \cr
\cr
&&= \sum_{N=1}^{\infty} {1\over N!} \sum_{k=1}^N
\lt\{ \sum_{n_k=1}^{\infty}~ {j_{n_k}\over \sqrt{2n_k}} {1\over \sqrt{2m}} ~
\oint ~ {dz\over 2\pi} {dz_k\over 2\pi}~ (\tan z)^{-m}~ (\tan z_k)^{-n_k} 
T_2(z,z_k,t_0) \rt\} \cr
\cr
&&\lt\{ \sum_{\{n_i\}_{i\neq k}} 
\prod_{i\neq k}^N \lt( {j_{n_i} \over \sqrt{2n_i}}\rt)  
\oint ~\prod_{i\neq k}^N \lt[ \frac{dz_i}{2\pi}~ (\tan z_i)^{-n_i} \rt] 
~\L \prod_{i\neq k}^N \partial X(z_i)~ e^{ipX}(0) \R_{\sigma(t_0)} \rt\} \cr
\cr
&& +~ {ip\over \sqrt{2m}}~\lt(\oint~ {dz\over 2\pi}~ (\tan z)^{-m}~T_1(z,t_0) 
\rt)~ G(j, p, t_0),
\eea
where in the last line we have made use of eq.(\ref{G2}).
The summand of the $k$ summation has two parts which are kept in two pairs of
curly brackets in the first two lines. The $k$ dependence of the quantity in 
the first pair of curly brackets comes only through $n_k$ and $z_k$ both of 
which are summed over. Hence it is $k$ independent.  For the second part also 
the $k$ dependence is only formal, as we can just remove it by renaming $n_i$ 
by $n_{i-1}$ for $i\geq k$. Therefore finally we get,

\bea
\partial_{j_m}G(j, p, t_0) &=& \lt[ -\sum_{n=1}^{\infty}~ {j_n \over 
2\sqrt{mn}}~\oint ~{dz\over 2\pi i }{dw\over 2\pi i}~(\tan z)^{-m}~
(\tan w)^{-n}~T_2(z,w,t_0) \rt. \cr 
\cr
&&\lt.+~ {ip\over \sqrt{2m}}~\oint~ {dz\over 2\pi}~ (\tan z)^{-m}~
T_1(z,t_0) \rt]~ G(j, p, t_0) \cr 
\cr
&=& \lt[-\sum_{n\geq 1} \widehat Q_{mn}~j_n ~+~ p~\widetilde L_m \rt]~
G(j, p, t_0),  \label{dG3}
\eea
where, 
\bea
\widehat Q_{mn}(\epsilon) &=& {1\over 2\sqrt{mn}} \oint {dz \over 2\pi i} 
\oint {dw \over 2\pi i}~ (\tan z)^{-m}~ (\tan w)^{-n}~ \partial_z \partial_w
{\cal G}^N(z,w,t_0) ,\cr
\cr
\widetilde L_n(\epsilon) &=& -{1\over \sqrt{2n}} \oint {dz\over 2\pi i}~ 
(\tan z)^{-n}~ \partial_z {\cal G}^N(z, 0, t_0).
\label{QhatLtilde}
\eea
Note that we have used the definitions (\ref{T2T1}).
$\widehat Q_{mn}$ and $\widetilde L_n$ are considered to be functions of
$\epsilon$ as $t_0=\pi /4 + \epsilon $. Now integrating eq.(\ref{dG3}) for all
values of $m$ we get,
\bea
G(j, p, t_0) = G(0, p, t_0)~\exp \lt(-~{1\over 2}~ 
j.\widehat Q(\epsilon).j~+~p~ j.\widetilde L(\epsilon) \rt), \label{G3}
\eea 
where $G(0, p, t_0)=\L e^{ipX}(0) \R_{\sigma(t_0)}$ is the constant of 
integration. 
Now using eqs.(\ref{Xip_X2}) and (\ref{G3}) one can express 
$|\Xi^{\prime}_X(\epsilon ) \R$ in terms of ${\cal G}^N$ and 
$\L e^{ipX}(0) \R_{\sigma(t_0)}$:
\bea
|\Xi^{\prime}_X(\epsilon)\R &=& \oint~ dp ~~ \widehat{h}(p,\epsilon) ~ \exp 
\lt(-~{1\over 2}~ a^{\dagger}.\widehat Q(\epsilon).a^{\dagger} ~+~ p~ 
\widehat L(\epsilon).a^{\dagger} \rt)~ |p\R,  \cr
\cr
\widehat L_n(\epsilon) &=& (-1)^{n}~\widetilde L_n(\epsilon), \cr
\cr
\widehat{h}(p,\epsilon) &=& {\cal{N}}~(2\epsilon )^{2h}~ 
\L e^{-ipX}(0) \R_{\sigma(t_0)}.  \label{Xip_X3}
\eea
To complete the derivation one needs to know the expressions for 
${\cal G}^N(z,w,t_0)$ and $\L e^{ipX}(0) \R_{\sigma(t_0)}$. Borrowing the 
results from appendix \ref{sigma} (eqs.(\ref{T2}, \ref{T1}, \ref{e2})) we get,
\bea
\widehat Q_{mn}(\epsilon) &=& {1\over \sqrt{mn}} \oint {dz \over 2\pi i} 
\oint {dw \over 2\pi i}~ (\tan z)^{-m}~ (\tan w)^{-n}~ (z-w)^{-2}  \cr
\cr
&&~~~~~~~~~~~~~~~~~~~~~~~~~(t_0^2-z^2)^{-1/2} (t_0^2-w^2)^{-1/2} (zw - t_0^2),
\cr
\cr
\widehat L_n(\epsilon) &=& (-1)^{n} \sqrt{2\over n}~t_0 ~\oint 
{dz\over 2\pi i}~ 
(\tan z)^{-n}~z^{-1} ~(t_0^2 - z^2)^{-1/2}, \cr
\cr
\widehat h(p,\epsilon) &=& {\cal{N}}~\lt(1+{\pi \over 4\epsilon}\rt)^{-1/8}~ 
\lt(2\epsilon + {\pi \over 2}\rt)^{-p^2}.
\label{QhatLhathhat}
\eea

Therefore $|\Xi^{\prime}_X(\epsilon)\R$ has the same form as 
$|\Psi^{\prime}_X(b)\R$ (eq.(\ref{Psip_X2})). It is also clear that 
$\epsilon$ plays the role similar to $b$. Now the question is 
how these two states are related. Surprisingly, we find numerical evidence 
that $|\Xi^{\prime}_X(\epsilon)\R$ is the same one parameter family of states 
as $|\Psi^{\prime}_X(b)\R$ with $\epsilon$ being a suitable function of $b$. 
We will present these results in the next section.

Before ending this section we will briefly indicate how one derives the 
expressions in eqs. (\ref{sliver-oscillator}) and (\ref{S_mn}) using the method
 described above. The definition (\ref{sliver-m-defined}) can be considered 
to be the special case of eq.(\ref{sliverp-m-defined}) where 
$\sigma^{\pm}=$ identity and $h=0$. Using this and following the similar
derivation given in appendix \ref{DXip_X2} it is straightforward to show that 
the analog of eq.(\ref{Xip_X2}) reads,
\bea
|\Xi_X\R = \widehat {\cal N} ~\int dp~ G(j_n~\rightarrow~ (-1)^{n+1}
a_n^{\dagger}, -p)~|p\R ,  \label{Xi_X}
\eea
where,
\bea
G(j, p) \equiv \L f \circ \exp(j.a^{\dagger}) ~ e^{ipX}(0) \R.
\eea
Clearly the correlation function has to be computed in the BCFT of D25-brane. 
Proceeding in the same way as discussed in this section one gets,
\bea
G(j,p) = \delta (p) ~\exp \lt(-{1\over 2} ~j.\widehat S.j \rt),
\eea
where $\widehat S_{mn}$'s are given by eq.(\ref{S_mn}) which is same as
$\widehat Q_{mn}$ given in eqs.(\ref{QhatLtilde}) with 
$\partial_z \partial_w {\cal G}^N(z,w,t_0)$ replaced by $\L\partial X(z)~
\partial X(w) \R = -2 (z-w)^{-2}$ (see eq.(\ref{dX-dX})). This result along
with eq.(\ref{Xi_X}) and the fact that with BCFT corresponding to the 
D25-brane all the directions should look the same, immediately justifies 
eq.(\ref{sliver-oscillator}). 
%
%
\sectiono{Numerical results}
\label{numerics}
To show that $|\Xi^{\prime}_X(\epsilon)\R$ and $|\Psi^{\prime}_X(b)\R$ are 
actually the same family of solutions we need to establish that the following 
relations are true with some suitable function $\epsilon (b)$:
\bea
\widehat Q_{mn}(\epsilon) &=& Q_{mn}(b), ~~~~~~~~~~~~\forall m,n\geq 1,
\label{checkQ} \\
\widehat L_n(\epsilon) &=& L_n(b),~~~~~~~~~~~~~~\forall n\geq 1, 
\label{checkL} \\
\widehat h(p, \epsilon ) &=& h(p,b).
\label{checkh}
\eea
In the following we will perform partial numerical verification of the above 
equations.

\subsectiono{Determination of $\epsilon (b)$}
To verify eqs. (\ref{checkQ}) and (\ref{checkL}) one needs to know 
$\epsilon (b)$ which we determine by equating the widths of the gaussians 
$\widehat h(p, \epsilon )$ and $h(p,b)$ as functions of $p$. We get the 
following relation:
\bea
\epsilon = {1\over 2} \exp\lt[{b \over 2}\lt({1\over 2}+{S^{\prime}_{00}\over
1-S^{\prime}_{00}} \rt) \rt] ~-~ {\pi \over 4} \label{epsilon(b)}
\eea
$\epsilon $ in the definition (\ref{sliver-defined}), is a real positive 
parameter. Hence the two solutions (\ref{Psip_X2}) and (\ref{Xip_X3}) can 
coincide only in the range $0\leq \epsilon < \infty $. Numerically this range 
corresponds
%
%
\begin{figure}[ht]
\leavevmode
\begin{center}
\epsfbox{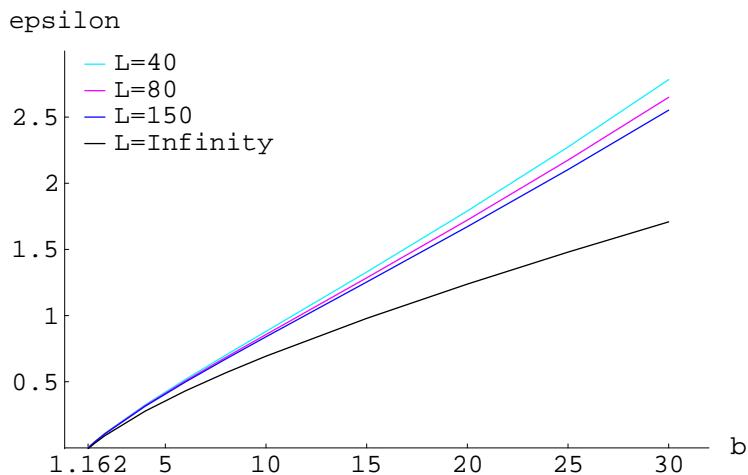}
\end{center}
\caption[]{Numerical plot of $\epsilon (b)$ for $L=40, ~ 80, ~150,
~ \infty $. The $L=\infty $ curve has been obtained by extrapolating the 
numerical data with the help of a fit of the form: 
$a_0 ~+~ a_1/\log (L) ~+~ a_2/(\log (L))^2 $. } 

\label{fig1}
\end{figure}
%
to $b_0\leq b<\infty $, where $b_0\equiv b(\epsilon = 0)$. We have performed 
numerical computations only up to $L=150$. At this level we find 
$\epsilon (b=1.16243) = 3.86646\times 10^{-7}$. We have displayed in 
fig.\ref{fig1},
the numerical plot of the function $\epsilon (b)$ up to $b=30$ for 
$L=40, ~80, ~150$ and $ \infty $.

\begin{figure}[!ht]
\leavevmode
\begin{center}
\epsfbox{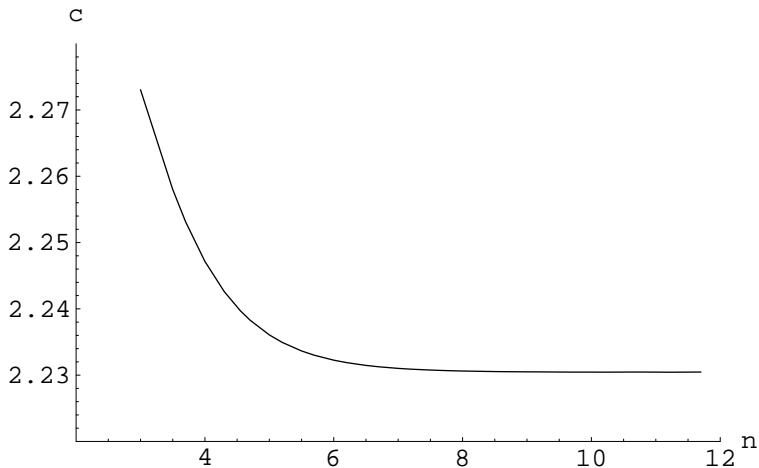}
\end{center}
\caption[]{Plot of $c=b^{1/2}(S^{\prime}_{00} + 1)$ against $n$, where 
$n=\log_{10}b$. $L=150$ result. } \label{fig3}
\end{figure}

It can be shown that $\epsilon $ diverges as $b$ goes to infinity. 
For this one uses the 
large $b$ dependence of $S^{\prime}_{00}$ (see eq.(\ref{epsilon(b)})). Using 
the expression for the matrix $S^{\prime}$ given in eqs.(\ref{NSTXCp}) and the
$b$ dependent matrices $V^{\prime rs}(b)$ given in ref.\cite{RSZ2} one can
show: $ \lim_{b\rightarrow \infty} S^{\prime}_{00} \rightarrow -1$.  
Numerically we find the following large $b$ behaviour of $S^{\prime}_{00}$
(see fig.\ref{fig3}):
\bea
S^{\prime}_{00} \sim -1 + c. b^{-1/2} + {\cal O}(b^{-1}), 
\eea
where $c$ is a positive constant, determined numerically to be around 2.23
(fig.\ref{fig3}) at $L=150$. Using this expansion in eq.(\ref{epsilon(b)})
it is easy to show:
\bea
\lim_{b\rightarrow \infty }\epsilon \rightarrow \infty
\eea
%
%
%
\subsectiono{Verifying eqs. (\ref{checkQ}) and (\ref{checkL})}
Using the functional relation (\ref{epsilon(b)}) we check eqs.(\ref{checkQ}) 
and (\ref{checkL})
for a few values of $m$ and $n$. We present these numerical results in tables
\ref{t-Q111}, \ref{t-Q112}, $\cdots $, \ref{t-L62}.  In these tables we 
have displayed the difference between l.h.s. and r.h.s. of eqs. (\ref{checkQ}) 
and (\ref{checkL})  computed in level truncation.  Here ``level truncation'' 
carries the usual meaning. We have performed the computations at a variety of 
values of $b$, namely $10,~15,~20$ and $25$ up to level $L=150$. At each level 
we 
have displayed the value of the percentage deviation $P_{mn}$ or $P_n$ defined 
through:
\bea
P_{mn} \equiv 100 \lt({\widehat Q_{mn} - Q_{mn} \over Q_{mn}}\rt) ,&& 
P_{n} \equiv 100 \lt({\widehat L_n - L_n \over L_n}\rt).
\eea
In the cases where magnitude of the percentage deviation is around $20$ or more
at $L=20$,
we have extrapolated the result to $L=\infty $ by using a fit of the form:
$$ a_0 ~+~ a_1 \lt({1\over \log L}\rt) ~+~ a_2 \lt({1\over \log L}\rt)^2. $$

The numerical results show that typically $P_{mn}$ or $P_n$ approaches 
monotonically to zero as $L$ increases when respectively 
$| \widehat Q_{mn} - Q_{mn} |$ or $| \widehat L_n - L_n | \geq 10^{-2}$ at 
$L=20$. The monotonic nature is absent when $| \widehat Q_{mn} - Q_{mn} |$ or 
$| \widehat L_n - L_n | \approx 10^{-3}$ at $L=20$. But in these cases
$(\widehat Q_{26}-Q_{26}) $ itself is very small so that the numerical values 
are not quite reliable. There is only one exceptional case, namely 
$Q_{24}$ at $b=10$, where although $|\widehat Q_{24} - Q_{24}| = 0.004610 $ at 
$L=20$, $P_{24}$ decreases monotonically as $L$ increases. It should be 
emphasized here 
that neither of $Q_{mn}$ and $\widehat Q_{mn}$ can be computed exactly.
$Q_{mn}$ is always computed in level truncation and here $\widehat Q_{mn}$ 
also depends\footnote{This is also true for $\widehat L_n$.} on the level as 
it uses the function $\epsilon (b)$ 
(eq.(\ref{epsilon(b)})) which is evaluated in level truncation. One can check
from fig.\ref{fig1} that $\epsilon (b)$ depends reasonably on $L$. 
\newpage
%
%
%
\begin{table}[ht]
\begin{center}\def\st{\vrule height 2.7ex width 0ex}
\begin{tabular}{|c|c|c|c|c|c|c|}  \hline \hline
\multicolumn{1}{|c|}{$b$} & \multicolumn{3}{|c|}{10} & 
\multicolumn{3}{|c|}{15} \st\\[1ex]
\hline \hline
$L$ & $Q_{11}$ & $\widehat Q_{11}-Q_{11}$ & $P_{11}$ & $Q_{11}$ & 
$\widehat Q_{11}-Q_{11}$ & $P_{11}$  \st\\[1ex]
\hline \hline
$20$  & $0.095904$ & $0.063869$ & $66.57$ &  0.14419 & 
$0.083102$ & $57.65$ \st\\[1ex]
\hline
$40$  & $0.097524$ &  0.05576 & 57.18 & 0.1467  
& 0.07485 & 51.02 \st\\[1ex]
\hline
$60$  & 0.098219 & 0.051846 & 52.79 & 0.14779 
& 0.070844 & 47.94 \st \\[1ex]
\hline
$80$  & 0.098626 & 0.049385 & 50.07 & 0.14844 & 0.06831 & 46.02  \st\\[1ex]
\hline
$100$ & 0.098901 & 0.04764 & 48.17 & 0.14888 & 0.066506 & 44.67  \st\\[1ex]
\hline
$150$ & 0.099324 & 0.044774 & 45.08 & 0.149558 & 0.063539 & 42.48 \st\\[1ex]
\hline
$\infty $ & 0.103999 & 0.004353 & 4.19 & 0.157412 & 0.021230 & 13.49 \st\\[1ex]
\hline \hline
\end{tabular}
\caption{Numerical check of eq.(\ref{checkQ}) for $(m,n)=(1,1)$ at 
$b=10,~15$ in level truncation.}  \label{t-Q111}
\end{center}
\end{table} 
%
%
\begin{table}[!ht]
\begin{center}\def\st{\vrule height 2.7ex width 0ex}
\begin{tabular}{|c|c|c|c|c|c|c|c|}  \hline \hline
\multicolumn{1}{|c|}{$b$} & \multicolumn{3}{|c|}{20} & 
\multicolumn{3}{|c|}{25}  \st\\[1ex]
\hline \hline
$L$ & $Q_{11}$ & $\widehat Q_{11}-Q_{11}$ & $P_{11}$ & $Q_{11}$ & 
$\widehat Q_{11}-Q_{11}$ & $P_{11}$  \st \\[1ex]
\hline \hline
$20$ &  0.17272 & 0.090202 & 52.22 & 0.191686 & 0.092387 & 48.2 \st \\[1ex] 
\hline
$40$  & 0.17592 & 0.082084 & 46.66 & 0.195448 & 0.084469 & 43.22 \st\\[1ex]
\hline
$60$  & 0.17733 & 0.078135 & 44.06 & 0.197114 & 0.080619 & 40.9 \st\\[1ex]
\hline
$80$  & 0.17817 & 0.075634 & 42.45 & 0.198111 & 0.07818 & 39.46 \st\\[1ex]
\hline
$100$  & 0.17875 & 0.073848 & 41.31 & 0.198794 & 0.076437 & 38.45 \st\\[1ex]
\hline
$150$  & 0.179641 & 0.070906 & 39.47 & 0.199866 & 0.073563 & 36.81 \st\\[1ex]
\hline
$\infty $ & 0.19031 & 0.004162 & 2.19 & 0.212775 & 0.001696 &  0.80 \st\\[1ex]
\hline \hline
\end{tabular}
\caption{Numerical check of eq.(\ref{checkQ}) for $(m,n)=(1,1)$ at 
$b=20,~25$ in level truncation.}  \label{t-Q112}
\end{center}
\end{table} 
\newpage
%
%
\begin{table}[ht]
\begin{center}\def\st{\vrule height 2.7ex width 0ex}
\begin{tabular}{|c|c|c|c|c|c|c|}  \hline \hline
\multicolumn{1}{|c|}{$b$} & \multicolumn{3}{|c|}{10} & 
\multicolumn{3}{|c|}{15} \st\\[1ex]
\hline \hline
$L$ & $Q_{13}$ & $\widehat Q_{13}-Q_{13}$ & $P_{13}$ & $Q_{13}$ & 
$\widehat Q_{13}-Q_{13}$ & $P_{13}$ \st\\[1ex]
\hline \hline
$20$ & -0.064555 & -0.015287 & 23.68 & -0.079509 & -0.022967 & 28.89 
\st\\[1ex] \hline
$40$ & -0.066134 & -0.011948 & 18.07 & -0.081656 & -0.018588 & 22.76 
\st\\[1ex] \hline
$60$ & -0.066822 & -0.010416 & 15.59 & -0.0826 & -0.016532 & 20.01 
\st\\[1ex]  \hline
$80$ & -0.067226 & -0.009481 & 14.1 & -0.083160 & -0.015261 & 18.35 
\st\\[1ex]  \hline
$100$ & -0.067499 & -0.008832 & 13.08 & -0.083541 & -0.014369 & 17.2 
\st\\[1ex]  \hline
$150$ & -0.067919 & -0.007798 & 11.48 & -0.084131 & -0.012931 & 15.37 
\st\\[1ex]   
\hline
$\infty $ & -0.072715 & 0.005613 & -7.72 & -0.072715 & 0.006507 & -8.95
\st\\[1ex]
\hline \hline
\end{tabular}
\caption{Numerical check of eq.(\ref{checkQ}) for $(m,n)=(1,3)$ at 
$b=10,~15$ in level truncation.} \label{t-Q131}
\end{center}
\end{table}
%
%
\begin{table}[!ht]
\begin{center}\def\st{\vrule height 2.7ex width 0ex}
\begin{tabular}{|c|c|c|c|c|c|c|}  \hline \hline
\multicolumn{1}{|c|}{$b$} & \multicolumn{3}{|c|}{20} & 
\multicolumn{3}{|c|}{25} \st\\[1ex]
\hline \hline
$L$ & $Q_{13}$ & $\widehat Q_{13}-Q_{13}$ & $P_{13}$ & $Q_{13}$ & 
$\widehat Q_{13}-Q_{13}$ & $P_{13}$ \st\\[1ex]
\hline \hline
$20$ & -0.08850 & -0.029099 & 32.88 & -0.094546 & -0.033075 & 34.98 
\st\\[1ex]  \hline
$40$  & -0.091064 & -0.024321 & 26.71 & -0.097428 & -0.028163 & 28.91 
\st\\[1ex]  \hline
$60$  & -0.092202 & -0.022054 & 23.92 & -0.098716 & -0.025820 & 26.16 
\st\\[1ex]   \hline
$80$  & -0.092880 & -0.020643 & 22.22 & -0.099487 & -0.024358 & 24.48
\st\\[1ex]   \hline
$100$  & -0.093342 & -0.019648 & 21.05 & -0.100015 & -0.023324 & 23.32 
\st\\[1ex]  \hline
$150$  & -0.094064 & -0.018035 & 19.17 & -0.100841 & -0.021641 & 21.46 
\st\\[1ex]
\hline
$\infty $ & -0.102734 & 0.004162 & -4.05 & -0.110942 & 0.001696 & -1.53
\st\\[1ex]
\hline 
\hline
\end{tabular}
\caption{Numerical check of eq.(\ref{checkQ}) for $(m,n)=(1,3)$ at 
$b=20,~25$ in level truncation.} \label{tt-Q132}
\end{center}
\end{table}
\newpage
%
%
\begin{table}[ht]
\begin{center}\def\st{\vrule height 2.7ex width 0ex}
\begin{tabular}{|c|c|c|c|c|c|c|}  \hline \hline
\multicolumn{1}{|c|}{$b$} & \multicolumn{3}{|c|}{10} & 
\multicolumn{3}{|c|}{15}  \st\\[1ex] 
\hline \hline
$L$ & $Q_{15}$ & $\widehat Q_{15}-Q_{15}$ & $P_{15}$ & $Q_{15}$ & 
$\widehat Q_{15}-Q_{15}$ & $P_{15}$ \st\\[1ex]
\hline \hline
$20$ & 0.044457 & 0.009871 & 22.20 & 0.053378 & 0.014552 & 27.26 
\st\\[1ex]  \hline
$40$ & 0.045886 & 0.007425 & 16.18 & 0.055269 & 0.011262 & 20.38 
\st\\[1ex]  \hline
$60$ & 0.046518 & 0.006305 & 13.55 & 0.056114 & 0.009726 & 17.33
 \st\\[1ex]  \hline
$80$ & 0.046892 & 0.005625 & 12.00 & 0.056617 & 0.008783 & 15.51 
\st\\[1ex]  \hline
$100$ & 0.047145 & 0.005155 & 10.93 & 0.056959 & 0.008125 & 14.26 
\st\\[1ex] \hline 
$150$ & 0.047535 & 0.004410 & 9.28 & 0.057491 & 0.007071 & 12.30 
\st\\[1ex] 
\hline
$\infty $ & 0.052152 & -0.005177 & -9.93 & 0.063944 & -0.006995 & -10.94
\st\\[1ex]
\hline
\hline
\end{tabular}
\caption{Numerical check of eq.(\ref{checkQ}) for $(m,n)=(1,5)$
 at $b=10,~15$ in level truncation.} \label{t-Q151}
\end{center}
\end{table}
%
%
\begin{table}[!ht]
\begin{center}\def\st{\vrule height 2.7ex width 0ex}
\begin{tabular}{|c|c|c|c|c|c|c|}  \hline \hline
\multicolumn{1}{|c|}{$b$} & \multicolumn{3}{|c|}{20} & 
\multicolumn{3}{|c|}{25}  \st\\[1ex]
\hline \hline
$L$ & $Q_{15}$ & $\widehat Q_{15}-Q_{15}$ & $P_{15}$ & $Q_{15}$ & 
$\widehat Q_{15}-Q_{15}$ & $P_{15}$ \st\\[1ex]
\hline \hline
$20$  & 0.058778 & 0.018997 & 32.32 & 0.062426 & 0.022237 & 35.62 
\st\\[1ex] \hline
$40$  & 0.061002 & 0.015290 & 25.06 & 0.064899 & 0.018345 & 28.27
\st\\[1ex] \hline
$60$  & 0.062003 & 0.013539 & 21.84 & 0.066019 & 0.016493 & 24.98 
\st\\[1ex] \hline
$80$  & 0.062601 & 0.012455 & 19.90 & 0.066692 & 0.015342 & 23.00 
\st\\[1ex] \hline
$100$  & 0.063010 & 0.011695 & 18.56 & 0.067153 & 0.014531 & 21.64 
\st\\[1ex] \hline 
$150$  & 0.063649 & 0.010469 & 16.45 & 0.067875 & 0.013220 & 19.48 
\st\\[1ex] 
\hline
$\infty $ & 0.071531 & -0.006240 & -8.72 & 0.076923 & -0.004874 & -6.34
\st\\[1ex]
\hline
\hline
\end{tabular}
\caption{Numerical check of eq.(\ref{checkQ}) for $(m,n)=(1,5)$ 
at $b=20,~25$ in level truncation.} \label{t-Q152}
\end{center}
\end{table}
\newpage
%
%
\begin{table}[ht]
\begin{center}\def\st{\vrule height 3ex width 0ex}
\begin{tabular}{|c|c|c|c|c|c|c|}  \hline \hline
\multicolumn{1}{|c|}{$b$} & \multicolumn{3}{|c|}{10} & 
\multicolumn{3}{|c|}{15} \st\\[1ex]
\hline \hline
$L$ & $Q_{22}$ & $\widehat Q_{22}-Q_{22}$ & $P_{22}$ & $Q_{22}$ & 
$\widehat Q_{22}-Q_{22}$ & $P_{22}$ \st\\[1ex]
\hline \hline
$20$ & -0.079754 & -0.001975 & 2.48 & -0.075862 & 0.003572 & -4.71  
\st\\[1ex] \hline
$40$ & -0.082366 & -0.000510 & 0.61 & -0.078173 & 0.005258 & -6.73 
\st\\[1ex] \hline
$60$ & -0.083606 & 0.000145 & 0.17 & -0.079263 & 0.006018 & -7.59 
\st\\[1ex]  \hline
$80$ & -0.084373 & 0.000534 & -0.63 & -0.079936 & 0.006473 & -8.10
\st\\[1ex]  \hline
$100$ & -0.084912 & 0.000798 & -0.94 & -0.080406 & 0.006784 & -8.44  
\st\\[1ex]  \hline
$150$ & -0.085779 & 0.001207 & -1.41 & -0.081162 & 0.007267 & -8.95 
\st\\[1ex]  
\hline
\hline
\end{tabular}
\caption{Numerical check of eq.(\ref{checkQ}) for $(m,n)=(2,2)$ 
at $b=10,~15$ in level truncation.} \label{t-Q221}
\end{center}
\end{table}
%
%
\begin{table}[!ht]
\begin{center}\def\st{\vrule height 3ex width 0ex}
\begin{tabular}{|c|c|c|c|c|c|c|}  \hline \hline
\multicolumn{1}{|c|}{$b$} & \multicolumn{3}{|c|}{20} & 
\multicolumn{3}{|c|}{25}   \st\\[1ex]
\hline \hline
$L$ & $Q_{22}$ & $\widehat Q_{22}-Q_{22}$ & $P_{22}$ & $Q_{22}$ & 
$\widehat Q_{22}-Q_{22}$ & $P_{22}$ \st\\[1ex]
\hline \hline
$20$  & -0.073516 & 0.004370 & -5.94 & -0.071926 &  0.004047 & -5.63
\st\\[1ex]  \hline
$40$ & -0.075639 & 0.006135 & -8.11 & -0.073918 & 0.005825 & -7.88  
\st\\[1ex] \hline
$60$ & -0.076635 & 0.006937 & -9.05 & -0.074848 & 0.006636 & -8.87 
\st\\[1ex] \hline
$80$ & -0.077247 & 0.007419 & -9.60 & -0.075419 & 0.007125 & -9.45 
\st\\[1ex]  \hline
$100$  & -0.077675 & 0.007749 & -9.98 & -0.075817 & 0.007462 & -9.84  
\st\\[1ex] \hline
$150$  & -0.078360 & 0.008267 & -10.55 & -0.076452 & 0.007991 & -10.45 
\st\\[1ex] 
\hline \hline
\end{tabular}
\caption{Numerical check of eq.(\ref{checkQ}) for $(m,n)=(2,2)$ 
at $b=20,~25$ in level truncation.} \label{t-Q222}
\end{center}
\end{table}
\newpage
%
%
%
\begin{table}[ht]
\begin{center}\def\st{\vrule height 3ex width 0ex}
\begin{tabular}{|c|c|c|c|c|c|c|}  \hline \hline
\multicolumn{1}{|c|}{$b$} & \multicolumn{3}{|c|}{10} & 
\multicolumn{3}{|c|}{15} \st\\[1ex]
\hline \hline
$L$ & $Q_{24}$ & $\widehat Q_{24}-Q_{24}$ & $P_{24}$ & $Q_{24}$ & 
$\widehat Q_{24}-Q_{24}$ & $P_{24}$ \st\\[1ex]
\hline \hline
$20$ & 0.053782 & 0.004610 & 8.57 & 0.051590 & 0.000756 & 1.47 
\st\\[1ex] \hline
$40$ & 0.055938 & 0.003106 & 5.55 & 0.053533 & -0.000742 & -1.39   
\st\\[1ex] \hline
$60$ & 0.056966 & 0.002403 & 4.22 & 0.054455 & -0.001432 & -2.63
\st\\[1ex]  \hline
$80$ & 0.057602 & 0.001976 & 3.43 & 0.055023 & -0.001848 & -3.36 
\st\\[1ex]  \hline
$100$ & 0.058048 & 0.001681 & 2.90 & 0.055420 & -0.002133 & -3.85
\st\\[1ex]  \hline
$150$ & 0.058763 & 0.001215 & 2.07 & 0.056055 & -0.002581 & -4.60  
\st\\[1ex]  
\hline
\hline
\end{tabular}
\caption{Numerical check of eq.(\ref{checkQ}) for $(m,n)=(2,4)$ 
at $b=10,~15$ in level truncation.} \label{t-Q241}
\end{center}
\end{table}
%
%
\begin{table}[!ht]
\begin{center}\def\st{\vrule height 3ex width 0ex}
\begin{tabular}{|c|c|c|c|c|c|c|}  \hline \hline
\multicolumn{1}{|c|}{$b$} & \multicolumn{3}{|c|}{20} & 
\multicolumn{3}{|c|}{25}   \st\\[1ex]
\hline \hline
$L$ & $Q_{24}$ & $\widehat Q_{24}-Q_{24}$ & $P_{24}$ & $Q_{24}$ & 
$\widehat Q_{24}-Q_{24}$ & $P_{24}$ \st\\[1ex]
\hline \hline
$20$ & 0.050262 & -0.000283 & -0.56 & 0.049358 & -0.000410 & -0.83
\st\\[1ex]  \hline
$40$ & 0.052071 & -0.001809 & -3.47 & 0.051073 & -0.001947 & -3.81
\st\\[1ex] \hline
$60$ & 0.052925 & -0.002512 & -4.75 & 0.051880 & -0.002655 & -5.12 
\st\\[1ex] \hline
$80$ & 0.053450 & -0.002935 & -5.49 & 0.052374 & -0.003083 & -5.89 
\st\\[1ex]  \hline
$100$ & 0.053815 & -0.003226 & -5.99 & 0.052718 & -0.003376 & -6.40 
\st\\[1ex] \hline
$150$ & 0.054399 & -0.003681 & -6.77 & 0.053265 & -0.003837 & -7.20 
\st\\[1ex] 
\hline \hline
\end{tabular}
\caption{Numerical check of eq.(\ref{checkQ}) for $(m,n)=(2,4)$ 
at $b=20,~25$ in level truncation.} \label{t-Q242}
\end{center}
\end{table}
\newpage
%
%
\begin{table}[ht]
\begin{center}\def\st{\vrule height 3ex width 0ex}
\begin{tabular}{|c|c|c|c|c|c|c|}  \hline \hline
\multicolumn{1}{|c|}{$b$} & \multicolumn{3}{|c|}{10} & 
\multicolumn{3}{|c|}{15} \st\\[1ex]
\hline \hline
$L$ & $Q_{26}$ & $\widehat Q_{26}-Q_{26}$ & $P_{26}$ & $Q_{26}$ & 
$\widehat Q_{26}-Q_{26}$ & $P_{26}$ \st\\[1ex]
\hline \hline
$20$ & -0.040383 & -0.005042 & 12.48 & -0.038843 & -0.002041 & 5.25 
\st\\[1ex] \hline
$40$ & -0.042280 & -0.003613 & 8.54 & -0.040566 & -0.000666 & 1.64 
\st\\[1ex] \hline
$60$ & -0.043194 & -0.002932 & 6.79 & -0.041393 & -0.000021 & 00.05 
\st\\[1ex]  \hline
$80$ & -0.043761 & -0.002514 & 5.74 & -0.041904 & 0.000372 & -00.89 
\st\\[1ex]  \hline
$100$ & -0.044159 & -0.002224 & 5.04 & -0.042261 & 0.000643 & -1.52 
\st\\[1ex]  \hline
$150$ & -0.044796 & -0.001764 & 3.94 & -0.042832 & 0.001068 & -2.49 
\st\\[1ex]  
\hline
\hline
\end{tabular}
\caption{Numerical check of eq.(\ref{checkQ}) for $(m,n)=(2,6)$ 
at $b=10,~15$ in level truncation.} \label{t-Q261}
\end{center}
\end{table}
%
%
\begin{table}[!ht]
\begin{center}\def\st{\vrule height 3ex width 0ex}
\begin{tabular}{|c|c|c|c|c|c|c|}  \hline \hline
\multicolumn{1}{|c|}{$b$} & \multicolumn{3}{|c|}{20} & 
\multicolumn{3}{|c|}{25}   \st\\[1ex]
\hline \hline
$L$ & $Q_{26}$ & $\widehat Q_{26}-Q_{26}$ & $P_{26}$ & $Q_{26}$ & 
$\widehat Q_{26}-Q_{26}$ & $P_{26}$ \st\\[1ex]
\hline \hline
$20$  & -0.037906 & -0.001058 & 2.79 & -0.037268 & -0.000820 & 2.20 
\st\\[1ex]  \hline
$40$ & -0.039521 & 0.000322 & -8.15 & -0.038806 & 0.000564 & -1.45 
\st\\[1ex] \hline
$60$ & -0.040292 & 0.000967 & -2.40 & -0.039537 & 0.001211 & -3.06 
\st\\[1ex] \hline
$80$ & -0.040766 & 0.001359 & -3.33 & -0.039987 & 0.001603 & -4.01 
\st\\[1ex]  \hline
$100$  & -0.041098 & 0.001628 & -3.96 & -0.040300 & 0.001874 & -4.65 
\st\\[1ex] \hline
$150$  & -0.041626 & 0.002051 & -4.93 & -0.040798 & 0.002298 & -5.63 
\st\\[1ex] 
\hline \hline
\end{tabular}
\caption{Numerical check of eq.(\ref{checkQ}) for $(m,n)=(2,6)$ 
at $b=20,~25$ in level truncation.} \label{t-Q262}
\end{center}
\end{table}
\newpage
%
%
\begin{table}[ht]
\begin{center}\def\st{\vrule height 2.7ex width 0ex}
\begin{tabular}{|c|c|c|c|c|c|c|} \hline \hline
\multicolumn{1}{|c|}{$b$} & \multicolumn{3}{|c|}{10} & 
\multicolumn{3}{|c|}{15}   \st\\[1ex]
\hline \hline
$L$ & $L_2$ & $\widehat L_2-L_2$ & $P_2$ &  $L_2$ & 
$\widehat L_2-L_2$ & $P_2$  \st\\[1ex]
\hline \hline
$20$ & -0.364194 & -0.128913 & 35.40 & -0.392023 & -0.168603 & 43.01
 \st\\[1ex]  \hline
$40$ & -0.373869 & -0.112747 & 30.16 & -0.402758 & -0.152126 & 37.77 
\st\\[1ex]  \hline
$60$ & -0.378499 & -0.1049 & 27.72 & -0.407913 & -0.144056 & 35.32 
\st\\[1ex]  \hline
$80$ & -0.381390 & -0.099956 & 26.21 & -0.41114 & -0.138943 & 33.80 \st\\[1ex]
\hline
$100$ & -0.383433 & -0.096438 & 25.15 & -0.413425 & -0.135291 & 32.72 
\st\\[1ex] \hline
150 & -0.386752 & -0.090679 & 23.45 & -0.417147 & -0.129283 & 30.99 \st\\[1ex] 
\hline
$\infty $ & -0.432994 & -0.008689 & 2.01 & -0.469289 & -0.042665 & 9.09 \st 
\\[1ex]
\hline \hline
\end{tabular}
\caption{Numerical check of eq.(\ref{checkL}) for $n=2$ 
at $b=10,~15$ in level truncation.} \label{t-L21}
\end{center}
\end{table}
%
%
\begin{table}[!ht]
\begin{center}\def\st{\vrule height 2.7ex width 0ex}
\begin{tabular}{|c|c|c|c|c|c|c|} \hline \hline
\multicolumn{1}{|c|}{$b$} & \multicolumn{3}{|c|}{20} & 
\multicolumn{3}{|c|}{25}   \st\\[1ex]
\hline \hline
$L$ & $L_2$ & $\widehat L_2-L_2$ & $P_2$ & $L_2$ & 
$\widehat L_2-L_2$ & $P_2$ \st\\[1ex]
\hline \hline
$20$ & -0.410521 & -0.185730 & 45.25 & -0.424065 & -0.193342 & 45.59 
\st\\[1ex] \hline
$40$ & -0.421957 & -0.169384 & 40.14 & -0.436008 & -0.177242 & 40.65 
\st\\[1ex] \hline
$60$ & -0.427461 & -0.161341 & 37.74 & -0.441764 & -0.169302 & 38.32 
\st\\[1ex] \hline
$80$ & -0.430911 & -0.156229 & 36.26 & -0.445377 & -0.164248 & 36.88 
\st\\[1ex]  \hline
$100$ & -0.433357 & -0.15257 & 35.21 & -0.447939 & -0.160625 & 35.86 
\st\\[1ex]  \hline
150 & -0.437346 & -0.146534 & 33.50 & -0.452121 & -0.154641 & 34.20 \st\\[1ex]  
\hline
$\infty $ & -0.493419 & -0.058952 & 11.95 & -0.511047 & -0.067541 & 13.22 
\st\\[1ex]
\hline \hline
\end{tabular}
\caption{Numerical check of eq.(\ref{checkL}) for $n=2$ 
at $b=20,~25$ in level truncation.} \label{t-L22}
\end{center}
\end{table}
\newpage
%
%
\begin{table}[ht]
\begin{center}\def\st{\vrule height 2.7ex width 0ex}
\begin{tabular}{|c|c|c|c|c|c|c|}  \hline \hline
\multicolumn{1}{|c|}{$b$} & \multicolumn{3}{|c|}{10} & 
\multicolumn{3}{|c|}{15} \st\\[1ex]
\hline \hline
$L$ & $L_4$ & $\widehat L_4-L_4$ & $P_4$ & $L_4$ & 
$\widehat L_4-L_4$ & $P_4$ \st \\[1ex]
\hline \hline
$20$ & 0.197429 & 0.079438 & 40.24 & 0.213512 & 0.106990 & 50.11
\st\\[1ex] \hline
$40$ & 0.205318 & 0.067864 & 33.05 & 0.222368 & 0.094046 & 42.29 
\st\\[1ex] \hline
$60$ & 0.209112 & 0.062277 & 29.78 & 0.226643 & 0.087723 & 38.71 
\st\\[1ex]  \hline
$80$ & 0.211479 & 0.058776 & 27.79 & 0.229317 & 0.083734 & 36.51 
\st\\[1ex] \hline
$100$ & 0.213150 & 0.056297 & 26.41 & 0.231207 & 0.080895 & 34.99 
\st\\[1ex] \hline
150 & 0.215855 & 0.052265 & 24.21 & 0.234277 & 0.076248 & 32.55 \st\\[1ex] 
\hline
$\infty $ &  0.253767 & - 0.004626 & - 1.82 & 0.277538 & 0.009568 & 3.45 
\st\\[1ex]
\hline \hline
\end{tabular}
\caption{Numerical check of eq.(\ref{checkL}) for $n=4$ 
at $b=10,~15$ in level truncation.} \label{t-L41}
\end{center}
\end{table}
%
%
\begin{table}[!ht]
\begin{center}\def\st{\vrule height 2.7ex width 0ex}
\begin{tabular}{|c|c|c|c|c|c|c|}  \hline \hline
\multicolumn{1}{|c|}{$b$} & \multicolumn{3}{|c|}{20} & 
\multicolumn{3}{|c|}{25}    \st \\[1ex]
\hline \hline
$L$ & $L_4$ & $\widehat L_4-L_4$ & $P_4$ & $L_4$ & 
$\widehat L_4-L_4$ & $P_4$ \st\\[1ex]
\hline \hline
$20$ & 0.224276 & 0.123145 & 54.91 & 0.232196 & 0.132486 & 57.06 
\st\\[1ex]  \hline
$40$ & 0.233791 & 0.109760 & 46.95 & 0.242196 & 0.119019 & 49.14
\st\\[1ex]  \hline
$60$ & 0.238393 & 0.103176 & 43.28 & 0.247042 & 0.112367 & 45.48
\st\\[1ex] \hline
$80$ & 0.241277 & 0.099004 & 41.03 & 0.250081 & 0.108141 & 43.24
\st\\[1ex] \hline
$100$ & 0.243318 & 0.096025 & 39.46 & 0.252233 & 0.105119 & 41.68
\st \\[1ex]  \hline
150 & 0.246636 & 0.091132 & 36.95 & 0.255736 & 0.100143 & 39.16 \st\\[1ex]  
\hline
$\infty $ & 0.293560 & 0.020229 & 6.89 & 0.305394 & 0.027642 & 9.05 
\st\\[1ex]
\hline \hline
\end{tabular}
\caption{Numerical check of eq.(\ref{checkL}) for $n=4$ 
at $b=20,~25$ in level truncation.} \label{t-L42}
\end{center}
\end{table}
\newpage
%
%
\begin{table}[ht]
\begin{center}\def\st{\vrule height 2.7ex width 0ex}
\begin{tabular}{|c|c|c|c|c|c|c|}  \hline \hline
\multicolumn{1}{|c|}{$b$} & \multicolumn{3}{|c|}{10} & 
\multicolumn{3}{|c|}{15}  \st\\[1ex]
\hline \hline
$L$ & $L_6$ & $\widehat L_6-L_6$ & $P_6$ & $L_6$ & 
$\widehat L_6-L_6$ & $P_6$ \st\\[1ex]
\hline \hline
$20$ & -0.135727 & -0.061953 & 45.65 & -0.147193 & -0.083386 & 56.65 
\st\\[1ex] \hline
$40$ & -0.142626 & -0.052404 & 36.74 & -0.154976 & -0.072390 & 46.71 
\st\\[1ex] \hline
$60$ & -0.145976 & -0.04777 & 32.72 & -0.158771 & -0.066998 & 42.20
\st\\[1ex] \hline
$80$ & -0.148072 & -0.044866 & 30.30 & -0.161150 & -0.063595 & 39.46 
\st\\[1ex] \hline
$100$ & -0.149551 & -0.042811 & 28.63 & -0.162832 & -0.061177 & 37.57 
\st\\[1ex] \hline
150 & -0.151945 & -0.039472 & 25.98 & -0.165562 & -0.057225 & 34.56 \st\\[1ex] 
\hline
$\infty $ & -0.18594 & 0.007933 & -4.27 & -0.204543 & -0.000255 & 0.12 
\st\\[1ex]
\hline \hline 
\end{tabular}
\caption{Numerical check of eq.(\ref{checkL}) for $n=6$ at $b=10,~15$ in 
level truncation.} \label{t-L61}
\end{center}
\end{table}
%
%
\begin{table}[!ht]
\begin{center}\def\st{\vrule height 2.7ex width 0ex}
\begin{tabular}{|c|c|c|c|c|c|c|}  \hline \hline
\multicolumn{1}{|c|}{$b$} & \multicolumn{3}{|c|}{20} & 
\multicolumn{3}{|c|}{25}   \st\\[1ex]
\hline \hline
$L$ & $L_6$ & $\widehat L_6-L_6$ & $P_6$ & $L_6$ & 
$\widehat L_6-L_6$ & $P_6$ \st\\[1ex]
\hline \hline
$20$ & -0.154897 & -0.097548 & 62.98 & -0.160580 & -0.106541 & 66.35 
\st\\[1ex]  \hline
$40$ & -0.163288 & -0.085937 & 52.63 & -0.169423 & -0.094710 & 55.90
\st\\[1ex] \hline
$60$ & -0.167387 & -0.080199 & 47.91 & -0.173751 & -0.088833 & 51.13 
\st\\[1ex] \hline
$80$  & -0.169962 & -0.076562 & 45.05 & -0.176472 & -0.085097 & 48.22 
\st\\[1ex] \hline
$100$  & -0.171785 & -0.073968 & 43.06 & -0.178400 & -0.082426 & 46.20 
\st\\[1ex]  \hline
150  & -0.174748 & -0.069714 & 39.89 & -0.181536 & -0.078036 & 42.99 
\st\\[1ex] 
\hline
$\infty $ & -0.217198 & -0.007736 & 3.56 & -0.226578 & -0.013637 & 6.02 
\st\\[1ex] 
\hline \hline 
\end{tabular}
\caption{Numerical check of eq.(\ref{checkL}) for $n=6$ at $b=20,~25$ in 
level truncation.}\label{t-L62}
\end{center}
\end{table}
%
\subsectiono{Verifying the normalization}
Now one also needs to verify that the normalizations of the gaussians 
$\widehat h(p, \epsilon )$ and $h(p,b)$ match, as we have not checked if 
$|\Xi^{\prime}_X(\epsilon)\R$ $*$-multiplies to itself separately in the 
oscillator language. The condition that the normalizations match reads,
\bea
R &\equiv& \sqrt{2\pi b\over 3}~(V^{rr}_{00} + b/2)^{-1}~
(1-S^{\prime}_{00})^{1/2}~\lt(1 + {\pi \over 4\epsilon} \rt)^{-1/8}
{\mbox{det}(1-X)^{1/2}\mbox{det}(1+T)^{1/2} \over 
\mbox{det}(1-X^{\prime})^{1/2}\mbox{det}(1+T^{\prime})^{1/2}}, \cr
&=& 1.  \label{checkN}
\eea
We present the numerical result for $(1-R)$ in table \ref{1-R}. Here also one
notices the same behaviour of the numerical data. $(1-R)$ approaches
monotonically to zero as $L$ increases when it is fairly large 
($|1-R|\geq 10^{-2}$). When $|1-R| \approx 10^{-3}$ as in the cases of $b=2$ 
and $6$, the behaviour is not that regular. 
%
\begin{center}\def\st{\vrule height 1.5ex width 0ex}
\begin{table}[ht]
\tabcolsep=1.4mm
\begin{tabular}{|c|c|c|c|c|c|c|c|c|c|} \hline \hline
~~~$b$ & 2 & 4 & 6 & 8 & 10 & 15 & 20 & 25 \\
$L$~~~ & &&&&&&& \\
\hline \hline
20~~~ & -0.003741 & -0.014667 & 0.003411 & 0.022327 & 0.039967 & 0.078096 & 
0.109525 & 0.136126 \st\\[1ex]
\hline
40~~~ & 0.002474 & -0.012070 & 0.003655 & 0.020720 &0.036840 & 0.072135 & 
0.101614 & 0.126813 \st\\[1ex]
\hline
60~~~ & 0.005389 & -0.010920 & 0.003643 & 0.019777 &0.035123 & 0.068950 & 
0.097396 & 0.121839 \st\\[1ex]
\hline
80~~~ & 0.007207 & -0.010216 & 0.003599 & 0.019129 & 0.033970 & 0.066828 & 
0.094586 & 0.118518 \st\\[1ex]
\hline
100~~~ & 0.008497 & -0.009723 & 0.003552 & 0.018643 & 0.033116 & 0.065260 & 
0.092506 & 0.116057 \st\\[1ex]
\hline
150~~~ & 0.019609 & -0.008923 & 0.003449 & 0.017802 & 0.031650 & 0.062577 &
0.088939 & 0.111827 \st\\[1ex]
\hline
$\infty $~~~ & 0.133653 & 0.000960 & 0.000114 & 0.003431 & 0.007829 & 0.019969 
& 0.032435 & 0.044717 \st\\[1ex]
\hline \hline
\end{tabular}
\caption{Numerical values of $(1-R)$ at various values of $b$ in level 
truncation. The $L=\infty $ results are obtained by extrapolation with a fit of
the following form: $ a_0 ~+~ a_1/\log (L) ~+~ a_2/(\log (L))^2 $.}  
\label{1-R} 
\end{table}
\end{center}
%
%
\sectiono{Generalization to Neumann-Dirichlet mixed boundary condition}
\label{generalization}
Our method of computing the oscillator expression of a D-brane solution in 
BCFT construction relied mainly on the validity of Wick's theorem. 
Applicability of Wick's theorem made it possible to get a simple first order
differential equation of $G(j,p,t_0)$ (eq.(\ref{dG3})) so that it can be 
integrated easily. Although the presence of $\sigma$ operators in the 
correlation function makes the boundary theory nontrivial, the final action 
is still quadratic as both the Neumann and Dirichlet boundary conditions 
connect the normal and tangential derivatives of $X$ on the world-sheet 
boundary in a linear way. In fact Wick's theorem is expected to be applicable 
in a more generic situation where the boundary condition involves the vanishing
 of an arbitrary
linear combination of the normal and tangential derivatives.
Let us consider the example of a D25-brane solution with a background constant
gauge  field strength $F_{\mu \nu}$ along the world volume in the VSFT on 
D25-brane. Here
$BCFT$ still corresponds to the usual D25-brane as before but now 
$BCFT^{\prime}$ corresponds to the following boundary condition ($z=e^{\tau + 
i\sigma}$):
\bea
\partial_{\sigma}X^{\mu}(\tau, \sigma) ~+~ \eta^{\mu \nu}~
\Omega_{\nu \gamma}(F)~
\partial_{\tau}X^{\gamma}(\tau, \sigma)|_{\sigma =0,\pi } = 0,  \label{ND}
\eea 
where $\Omega_{\mu \nu}(F)$ is an antisymmetric constant matrix proportional to
$F_{\mu \nu}$.
In this case one needs to compute the following normalized two point function 
(see discussion in appendix \ref{sigma}) with Neumann boundary condition on 
$|t|<t_0$ and boundary condition (\ref{ND}) on the rest of the real line. 
\bea
{\cal{G}}^{N\mu \nu}(z,w,t_0,F) = {\L X^{\mu}(z)~ X^{\nu}(w) \R_{\sigma(t_0)}
\over \L \sigma^+(t_0)~ \sigma^-(-t_0) \R },
\eea
where the $\sigma $ operators are defined in the same way as 
before with D.b.c. replaced by eq.(\ref{ND}). Just like the usual D25-brane 
the present solution should also not have any momentum excitation as the full 
translational invariance is still preserved. Therefore it is sufficient to 
consider $G(j,0,t_0)$ in this case. Using the same method that was followed in 
sec.\ref{D24-oscillator} 
one can show that the solution (denoted $|\Xi_F\R$)is given by the following:
\bea
|\Xi_F \R = \exp \lt(- {1\over 2} \sum_{m,n\geq 1} \widehat 
Q_{\mu \nu mn}(\epsilon,F)
~a^{\mu \dagger }_m a^{\nu \dagger }_n \rt) |0_{26}\R,
\eea
where,
\bea
\widehat Q_{\mu \nu mn}(\epsilon,F) = {1 \over 2\sqrt{mn}}
~\oint {dz \over 2\pi i}~{dw \over 2\pi i}~ (\tan z)^{-m}~ (\tan w)^{-n}
~\partial_z \partial_w {\cal{G}}^N_{\mu \nu}(z,w,t_0,F).
\eea
If we consider a lower dimensional brane solution then the transverse 
directions simply corresponds to D.b.c. and hence each of them looks the same 
as that found
in sec.\ref{D24-oscillator}.
%
\sectiono{Discussion}
\label{discussion}
\begin{enumerate}
\item
We have found numerical evidence that the two states
$|\Psi^{\prime}_X(b)\R$ (eq.(\ref{Psip_X2})) and
$|\Xi^{\prime}_X(\epsilon)\R$ (eqs.(\ref{Xip_X1})) are same with
 a specific relation (eq.(\ref{epsilon(b)})) between the respective gauge 
parameters $b$ and $\epsilon $ for a moderately large values of $b$, namely 
$10-25$. 

Here we will try to partially investigate the case of smaller values of $b$ . 
We find that
the numerical agreement is not good for smaller values of $b$ if we follow
the same method as that used in sec.\ref{numerics} to compare the two states.
For example, we find $P_{11}=75.9$ for $b=4$. It may be possible that the
two solutions are different for small values of $\epsilon $. But comparing the
two states in a different manner we again get good numerical agreement. In the
following we will elaborate on this issue.

Comparing the two solutions involve checking the conditions (\ref{checkQ}), 
(\ref{checkL}) and (\ref{checkh}) which give a set of infinite number of 
equations. One needs to choose only one of them to fix the function 
$\epsilon (b)$
and verify the rest of the infinite number of equations using that function.
The numerical result will certainly depend on the specific equation chosen to 
determine $\epsilon (b)$. In fact the numerical accuracy to which a specific
equation belonging to the set 
\{(\ref{checkQ}), (\ref{checkL}), (\ref{checkh})\} is satisfied for a certain 
value of $b$ will depend on this choice. In sec.\ref{numerics}, we have fixed 
the function $\epsilon (b)$ by equating (eq.(\ref{epsilon(b)})) the widths of 
the gaussians $\widehat h(p,\epsilon )$ and $h(p,b)$. It may happen that this
is not a good choice for $b<10$ at least for those equations which are not 
satisfied very well numerically, for example eq.(\ref{checkQ}) for 
$(m,n)=(1,1)$ (as it gives $P_{11}=75.9$). In that case some different choice 
should improve the numerical result. In the following we will give an 
example to demonstrate this.
\begin{figure}[ht]
\leavevmode
\begin{center}
\epsfbox{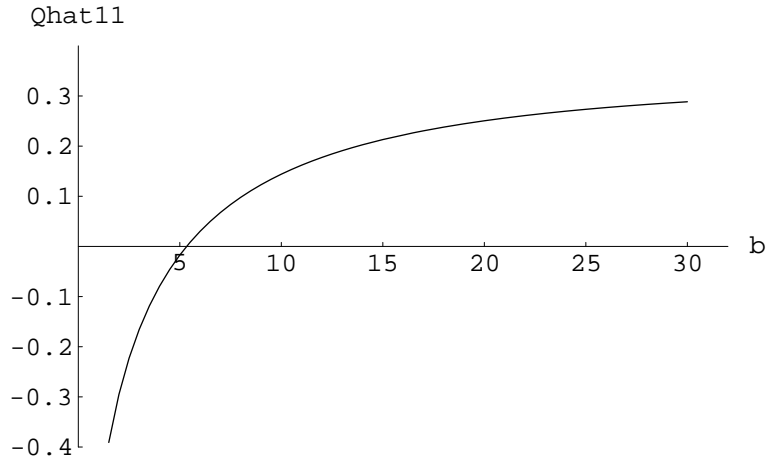}
\end{center}
\caption[]{Plot of $\widehat Q_{11}$ against $b$. $L=150$ result. } 
\label{Qhat11}
\end{figure}

In fig.\ref{Qhat11} we have plotted $\widehat Q_{11}$ against $b$ using the 
level 150 results.
 We see that $\widehat Q_{11}$ is reasonably flat for $b>10$.
The slope of the curve increases gradually below $b=10$ and it becomes 
considerably large below $b=5$. Since $\epsilon (b)$ is almost linear
(fig.\ref{fig1}) for such values of $b$ at $L=150$, the slope of 
$\widehat Q_{11}$ with respect to $\epsilon $ also becomes larger below $b=10$.
This means that small error in computing $\epsilon $ induces large error in 
the value of $\widehat Q_{11}$. This could be the possible reason for getting
unsatisfactory numerical results for small $b$. To check if it is true
one can perform the numerical verification in the opposite order, i.e. first
use the following equation to determine $t_0 = (\epsilon +\pi/4)$
(see eqs. (\ref{QLhB}) and (\ref{QhatLhathhat}))
as a function of $b$,
\bea
\widehat Q_{11}(\epsilon ) = Q_{11}(b),
\label{checkQ11}
\eea
then check if the following equation (condition for the equality of widths - 
eq.(\ref{epsilon(b)})) is satisfied:
\bea
t_0 = {1\over 2} \exp \lt[{b \over 2} \lt({1\over 2} ~+~ {S^{\prime}_{00} \over
1-S^{\prime}_{00}} \rt) \rt].
\label{t0}
\eea
We have performed computations for $b=2,~4,~6,~8$ and the numerical results 
seem to be much better in this case. We have presented these results in
tables \ref{t_01} and \ref{t_02} where we have used the following notations:
$t^Q_0$ is the value of $t_0$ obtained
by solving eq.(\ref{checkQ11}) and $t^{width}_0$ is the value of $t_0$
obtained by evaluating the r.h.s. of eq.(\ref{t0}). The percentage deviation 
$P$ has the following definition:
\bea
P = 100 \lt({t^Q_0 ~-~ t^{width}_0 \over t^Q_0}\rt)
\eea
The $L=\infty $ results are obtained by using the same fit as used earlier in 
this paper.
%
%
\begin{table}[ht]
\begin{center}\def\st{\vrule height 2.5ex width 0ex}
\begin{tabular}{|c|c|c|c|c|c|c|}  \hline \hline
\multicolumn{1}{|c|}{$b$} & \multicolumn{3}{|c|}{2} & 
\multicolumn{3}{|c|}{4}   \st\\[1ex]
\hline \hline
$L$ & $t^Q_0$ & $t^Q_0-t^{width}_0$ & $P$ & $t^Q_0$  & 
 $t^Q_0-t^{width}_0$ & $P$ \st\\[1ex]
\hline \hline
$20$ & 0.992418 & 0.093157 & 9.39 & 1.14835 & 0.02844 & 2.54
\st\\[1ex]  \hline
$40$ & 0.992437 & 0.096443 & 9.72 & 1.14885 & 0.03737 & 3.36
\st\\[1ex] \hline
$60$ & 0.992446 & 0.097959 & 9.87 & 1.14906 & 0.04152 & 3.75
\st\\[1ex] \hline
$80$ & 0.992452 & 0.098887 & 9.96 & 1.14918 & 0.04408 & 3.99
\st\\[1ex] \hline
$100$ & 0.992456 & 0.099534 & 10.03 & 1.14926 & 0.04586 & 4.16
\st\\[1ex]  \hline
150  & 0.992461 & 0.100568 & 10.13 & 1.14939 & 0.04874 & 4.43
\st\\[1ex] 
\hline
$\infty $ & 0.99254 & 0.114123 & 11.50 & 1.1507  & 0.086969 & 7.56
\st\\[1ex] 
\hline \hline 
\end{tabular}
\caption{Numerical results for $t^Q_0$ (solution of eq.(\ref{checkQ11}) and 
$t^{width}_0$ (r.h.s. of eq.(\ref{t0}) in level truncation at $b=2,~4$.}
\label{t_01}
\end{center}
\end{table}
%
%
\begin{table}[!ht]
\begin{center}\def\st{\vrule height 2.5ex width 0ex}
\begin{tabular}{|c|c|c|c|c|c|c|}  \hline \hline
\multicolumn{1}{|c|}{$b$} & \multicolumn{3}{|c|}{6} & 
\multicolumn{3}{|c|}{8}  \st\\[1ex]
\hline \hline
$L$ & $t^Q_0$ & $t^Q_0-t^{width}_0$ & $P$ & $t^Q_0$  & 
 $t^Q_0-t^{width}_0$ & $P$ \st\\[1ex]
\hline \hline
$20$ & 1.26718 & -0.05133 & -4.05 & 1.36587 & -0.14333 & -10.49 
\st\\[1ex] \hline
$40$ & 1.26873 & -0.03493 & -2.75 & 1.36895 & -0.1179 & -8.61 
\st\\[1ex] \hline
$60$ & 1.2694 & -0.02727 & -2.15 & 1.37027 & -0.10599 & -7.73
\st\\[1ex] \hline
$80$ & 1.26978 & -0.02255 & -1.78 & 1.37104 & -0.09862 & -7.19 
\st\\[1ex] \hline
$100$ & 1.27004 & -0.01923 & -1.51 & 1.37156 & -0.09343 & -6.81 
\st\\[1ex] \hline
140 & 1.27037 & -0.01473 & -1.16 & 1.37223 & -0.086398 & -6.30 \st\\[1ex] 
\hline
$\infty $ & 1.27472 & 0.058027 & 4.55 & 1.38116 & 0.028155 & 2.03
\st\\[1ex]
\hline \hline 
\end{tabular}
\caption{Numerical results for $t^Q_0$ (solution of eq.(\ref{checkQ11}) and 
$t^{width}_0$ (r.h.s. of eq.(\ref{t0}) in level truncation at $b=6,~8$.} 
\label{t_02}
\end{center}
\end{table}
\item
The numerical estimate of $b_0=b(\epsilon=0)$ also changes substantially
when one chooses different equations to determine $\epsilon (b)$. For example
using eq.(\ref{checkQ11}) we find: 
$\epsilon (b=1.16243) = 0.117631$,  
$\epsilon (b=0.395339) = 9.94763 \times 10^{-8}$. One can argue that $b_0$ has 
to be a nonzero positive number in the following way. Numerically ($L=150$) we 
find the following small $b$ behaviour of $S^{\prime}_{00}$:
\bea
S^{\prime}_{00} \sim 1 - 1.083 ~b^{1/2} + {\cal O}(b).
\eea
Using this in eq.(\ref{epsilon(b)}) one can show that, 
$\epsilon (0) = (1/2-\pi /4) <0$ which implies that $b_0$ has to be positive. 
\item
For $b=10-25$ one could try to fix $\epsilon (b)$ from the normalization 
(condition (\ref{checkN})) and verify if the widths match. This again is not
numerically efficient because of the following reason: the
$\epsilon $ dependence of eq.(\ref{checkN}) is a factor with a small exponent 
, namely $\lt(1+{\pi \over 4\epsilon } \rt)^{-1/8}$.
Therefore the other part of the equation which is computed numerically, has to 
be raised to the $8$th power to compute $\epsilon $. This increases the 
numerical error in the value of $\epsilon $ as compared to the one obtained 
from comparing the widths of the gaussians. Because of the same reason the 
numerical test of eq.(\ref{checkN}) is not very much sensitive to the 
functional relation (\ref{epsilon(b)}). 
\item
It is interesting that in some cases we get very small numbers even at a low 
level. For example we get $(\widehat Q_{26} - Q_{26}) = -0.003848$ at $L=6$ 
(the lowest value that can be chosen for $(m,n)=(2,6)$) and $b=25$, 
$(1-R) = -0.001445$ at $L=1$ and $b=2$. Typically in these cases we see that 
the quantity does not monotonically converge to zero as we increase level.
To understand this situation more one needs to perform computations at higher
levels.
\item
From table\ref{1-R} we see that the value of $(1-R)$ at a given level drifts 
away from zero as we increase $b$. This indicates the fact that at larger 
values of $b$ or $\epsilon $ the higher levels contribute more. This also 
happens when 
$\epsilon $ approaches zero as we can see it in the following table 
($L=150$ results):
%
\begin{center}\def\st{\vrule height 1.5ex width 0ex}
\begin{table}[h]
\tabcolsep=1.7mm
\begin{tabular}{|c|c|c|c|c|c|c|c|} \hline \hline
b & 1.7 & 1.2 & 1.167 & 1.16250 &  1.16243 \st\\[1ex] \hline
$\epsilon $ & 0.070290 & $5.242\times 10^{-3}$ & 
$6.412\times 10^{-4}$ & $1.02086\times 10^{-5}$ & $3.86646\times 10^{-7}$
\st\\[1ex] \hline
$1-R$ & 0.034139 & 0.258735 & 0.427176 & 0.658365 & 0.773086 \st\\[1ex]  
\hline \hline
\end{tabular}
\caption{$L=150$ results for $(1-R)$ as $\epsilon $ tends to zero.} 
\label{t-last}
\end{table}
\end{center}
%
\item
The method of translating the BCFT construction of a D-brane solution to 
its oscillator description as described in this paper only required a 
certain normalized two point function with specific boundary conditions on the 
real line. It will be interesting to explore the higher dimensional D-brane
solutions in the VSFT on a lower dimensional D-brane. For example one may try 
to see how the D25-brane solution looks in the VSFT on a D24-brane.

\end{enumerate}
\appendix
%
%
\sectiono{Proof of eq.(\ref{Xip_X2})}
\label{DXip_X2}
\setcounter{equation}{0}
Let us start from the definition of $|\Xi^{\prime}_X(\epsilon)\R $, namely 
eq.(\ref{Xip_X1}). Taking $|\phi \R = (a^{\dagger}_{n_1}~a^{\dagger}_{n_2}
\cdots )~|p\R $ one can write:
\bea
\begin{array}{l}
\L \Xi^{\prime}_X(\epsilon)|~(a^{\dagger}_{n_1}~a^{\dagger}_{n_2}
\cdots )~|p\R \\
\\
= {\cal{N}}~(2\epsilon )^{2h}~\L (f\circ a^{\dagger}_{n_1}~f\circ
a^{\dagger}_{n_2}\cdots )~ f\circ e^{ipX}(0)~\sigma^+(t_0) \sigma^-(-t_0)\R \\
\\
= {\cal{N}}~(2\epsilon )^{2h}~\L (f\circ a^{\dagger}_{n_1}~f\circ
a^{\dagger}_{n_2}\cdots )~ e^{ipX}(0)~\sigma^+(t_0) \sigma^-(-t_0)\R \\
\\
= {\cal{N}}~(2\epsilon )^{2h}~ \lt[( \partial_{j_{n_1}} \partial_{j_{n_2}} 
\cdots )~G(j,p,t_0)\rt]_{j=0} \\
\end{array}
\label{inner-derivative}
\eea
$|\phi \R $ considered above has the generic form of a state in 
${\cal{H}}_{BCFT_X}$. Let us now expand $|\Xi^{\prime}_X(\epsilon)\R$ in terms 
of the basis vectors of this type and take its BPZ conjugate: 
\bea
|\Xi^{\prime}_X(\epsilon)\R &=& \sum_{\{N_n\}}~ \int ~dp ~ C^{\Xi}_{\{N_n \}}
(p,\epsilon)~ \prod_{n=1}^{\infty}~\lt(a^{\dagger}_n \rt)^{N_n} ~|p\R, 
\label{Xip_X4} \\
\L\Xi^{\prime}_X(\epsilon)| &=& 
\sum_{\{N_n\}}~ \int ~dp ~ C^{\Xi}_{\{N_n \}}(p,\epsilon)~ 
(-1)^{\sum_n (1+n)N_n} \L p| \prod_{n=1}^{\infty}~\lt( a_n \rt)^{N_n},
\label{Xip_X5}
\eea
where, $N_n=0,1, \cdots \infty , \forall n\geq 1$. Therefore taking the BPZ 
inner product with the state
$\prod_{n=1}^{\infty}~\lt(a^{\dagger}_n \rt)^{N_n} ~|p\R $ one gets:
\bea
\L\Xi^{\prime}_X(\epsilon)|~\prod_{n=1}^{\infty}~\lt(a^{\dagger}_n \rt)^{N_n} 
~|p\R = C^{\Xi}_{\{N_n \}}(-p,\epsilon)~ (-1)^{\sum_n (1+n)N_n} 
\prod_{n=1}^{\infty} \lt( N_n ! \rt).
\label{inner}
\eea
The series expansion of $G(j,p,t_0)$
in powers of $j_n$'s will look like,
\bea
G(j,p,t_0) = \sum_{\{N_n\}} C^G_{\{N_n\}}(p,\epsilon )~ \prod_{n=1}^{\infty}~
(j_n)^{N_n}.
\eea
Therefore,
\bea
\lt[\prod_{n=1}^{\infty}\lt( \partial_{j_n}^{N_n}\rt) ~G(j,p,t_0) \rt]_{j=0} = 
C^G_{\{N_n\}}(p,\epsilon )~ \prod_{n=1}^{\infty}~\lt( N_n ! \rt).
\label{derivative}
\eea
Now using eqs.(\ref{inner-derivative}, \ref{inner}, \ref{derivative}) one can
write,
\bea
C^{\Xi}_{\{N_n\}}(p,\epsilon) = {\cal{N}}~(2\epsilon )^{2h}~ 
(-1)^{\sum_{n=1}^{\infty}(1+n)N_n}~ C^G_{\{N_n\}}(-p,\epsilon ).
\label{CC}
\eea
Using this expression in eq.(\ref{Xip_X4}) we get,
\bea
|\Xi^{\prime}_X(\epsilon )\R &=& {\cal{N}}~(2\epsilon )^{2h}~ \int ~dp~
\lt[~ \sum_{\{N_n\}}~C^G_{\{N_n\}}(-p,\epsilon )~ \prod_{n\geq 1}~
\lt( (-1)^{1+n} a^{\dagger}_n \rt)^{N_n} \rt] ~ |p\R \cr
&=& {\cal{N}}~(2\epsilon )^{2h}~ \int ~dp~
G(j_n\rightarrow (-1)^{1+n} a^{\dagger}_n, -p,t_0)~ |p\R ,
\eea
which is eq.(\ref{Xip_X2}).
%
%
\sectiono{Correlation functions with $\sigma^{\pm}(\pm t_0)$}
\label{sigma}
\setcounter{equation}{0}
As defined in subsection \ref{BCFT-construction}, $\sigma^{\pm}(t)$ (also 
called the twist operators in the literature) are the vertex operators of the 
ground states of strings connecting the Neumann and Dirichlet boundary 
conditions with opposite orientations. These are primaries with conformal
dimension $h=1/16$ and have the following OPE with $\partial X$ \cite{dpdq}:
\bea
\partial X(z)~ \sigma^{\pm}(w) = (z-w)^{-1/2}~ \tau^{\pm}(w) ~+~ 
\mbox{Reg.}, \label{dX-sigma} 
\eea
where $\tau^{\pm}(z)$ are the next excited twist fields. We will take the 
following normalization for $\sigma^{\pm}$:
\bea
\sigma^{\pm}(z)~\sigma^{\pm}(w) = (z-w)^{-1/8} ~+~ \mbox{Reg.}  
\label{sigma-sigma}
\eea
Eq.(\ref{adagger}) and the commutation relation 
$[a_m,a^{\dagger}_n]=\delta_{mn} $ fix the normalization of $\partial X$,
\bea
\partial X(z)~\partial X(w) = -2~(z-w)^{-2} ~+~ \mbox{Reg.}  \label{dX-dX}
\eea

Now we will use the above short distance behaviours and the method of 
analyticity to compute the twisted two point function
$g^{(2)} \equiv \L \partial X(z)~\partial X(w)~\sigma^+(u)~\sigma^-(v) \R $. 
First of all one notices that the singularities of $z$ as it approaches 
$w,~u$ and $v$ are $-2(z-w)^{-2}$, $(z-u)^{-1/2}$ and $(z-v)^{-1/2}$ 
respectively. Therefore $g^{(2)}$ should take the following form:
\bea
g^{(2)} = -2(z-w)^{-2}(z-u)^{-1/2}(z-v)^{-1/2}~ \lt[ f_1(w,u,v) ~+~ zf_2(w,u,v)
~+~ \cdots \rt].
\label{g^21}
\eea
But the large $z$ behaviour of $g^{(2)}$, namely $ \lim_{z\rightarrow \infty} 
z^2 g^{(2)}(z,w,u,v) \rightarrow $ finite, truncates the above series at the 
second term. Therefore we are left with computing the two unknown functions
$f_1(w,u,v)$ and $f_2(w,u,v)$. Now let us focus only on the $z\rightarrow w$
behaviour of $g^{(2)}$. Using eqs. (\ref{sigma-sigma}) and (\ref{dX-dX}) we
get,
\bea
g^{(2)} = -2(z-w)^{-2} ~ (u-v)^{-1/8} ~+~ \mbox{Reg.}
\label{g^22}
\eea 
Expanding eq.(\ref{g^21}) in powers of $(z-w)$ we get,
\bea
g^{(2)} &=& -2~(z-w)^{-2}\lt[(w-u)^{-1/2}(w-v)^{-1/2}(f_1 + wf_2) \rt] \cr
\cr
&& -2~(z-w)^{-1}\lt[(w-u)^{-1/2}(w-v)^{-1/2} \rt. \cr
\cr
&& \lt. \lt(f_2 ~-~ {1\over 2}(f_1 + wf_2) \lt[(w-u)^{-1} + 
(w-v)^{-1} \rt] \rt) \rt] ~+~ \mbox{Reg.}
\label{g^23}
\eea
Matching the coefficients of $(z-w)^{-2}$ and $(z-w)^{-1}$ from eqs. 
(\ref{g^22}) and (\ref{g^23}) we get the following two equations:
\bea
(f_1 + wf_2) &=& (w-u)^{1/2}(w-v)^{1/2} (u-v)^{-1/8}, \label{f1f21} \\
\cr
(f_1 + wf_2) &=& {2f_2\over (w-u)^{-1} + (w-v)^{-1} }. \label{f1f22}
\eea
Solving eqs. (\ref{f1f21}) and (\ref{f1f22}) for $f_1(w,u,v)$ and $f_2(w,u,v)$ 
we get,
\bea
f_1(w,u,v) &=& -(w-u)^{-1/2}(w-v)^{-1/2}(u-v)^{-1/8} \lt( {wu\over 2} ~+~ 
{wv\over 2} ~-~ uv \rt),  \label{f1} \\
\cr
f_2(w,u,v) &=& {1\over 2}(w-u)^{-1/2}(w-v)^{-1/2}(u-v)^{-1/8} ~(2w-u-v). 
\label{f2} 
\eea
Finally substituting this result in eq.(\ref{g^21}) we get,
\bea
\begin{array}{l}
\L \partial X(z)~\partial X(w)~\sigma^+(u)~\sigma^-(v) \R \\
\\
= -2~(z-w)^{-2}~(z-u)^{-1/2}~(z-v)^{-1/2}~(w-u)^{-1/2}~(w-v)^{-1/2} \\
\\
\;\;\;\;(u-v)^{-1/8} \lt( zw-\frac{1}{2}zu -\frac{1}{2}zv - \frac{1}{2}wu - 
\frac{1}{2}wv + uv \rt).
\end{array} \label{dX-dX-sigma-sigma}
\eea
The corresponding normalized correlator would be defined as,
\bea
\L \partial X(z)~\partial X(w)~\sigma^+(u)~\sigma^-(v) \R^N
\equiv {\L \partial X(z)~\partial X(w)~\sigma^+(u)~\sigma^-(v) \R \over
\L \sigma^+(u)~ \sigma^-(v) \R}.  \label{dX-dX-sigma-sigmaN}
\eea
Using eqs.(\ref{sigma-sigma}, \ref{dX-dX-sigma-sigma}, 
\ref{dX-dX-sigma-sigmaN}) we get,
\bea
T_2(z,w,t_0) &\equiv & \L \partial X(z)~\partial X(w)\R^N_{\sigma(t_0)} \cr
\cr
&=& {\L \partial X(z)~\partial X(w)~\sigma^+(t_0)~\sigma^-(-t_0)\R \over
\L\sigma^+(t_0)~\sigma^-(-t_0)\R} \cr 
\cr
&=& 2~(z-w)^{-2}~(t_0^2-z^2)^{-1/2}~(t_0^2-w^2)^{-1/2}~(zw-t_0^2).
\label{T2}
\eea
It is straightforward to verify that this is consistent with the following 
result for the normalized two point function of $X$ obtained by applying the 
method of images in ref.\cite{Hashi}:
\bea
{\cal{G}}^N(z,w,t_0) \equiv \L X(z)~X(w)\R_{\sigma(t_0)}^N = -2~\ln \lt[{1 ~-~ 
\sqrt{(z ~+~ t_0)(w ~-~ t_0) \over 
(w ~+~ t_0)(z ~-~ t_0)}
\over 1 ~+~ \sqrt{(z ~+~ t_0)(w ~-~ t_0) \over (w ~+~ t_0)(z ~-~ t_0)} } \rt].
\label{calG1}
\eea
Now we define,
\bea
T_1(z,t_0) \equiv \L \partial X(z) ~ X(0) \R^N_{\sigma(t_0)} =\partial_z 
{\cal{G}}^N(z,0,t_0) = -2~t_0~z^{-1}~(t_0^2 - z^2)^{-1/2}
\label{T1}
\eea  
Note that $T_2(z,w,t_0)$ and $T_1(z,t_0)$ defined above have been used in the 
first two equations of (\ref{QhatLhathhat}).

Now we are left with the job of computing $\L e^{ipX}(0) \R_{\sigma (t_0)}$
(the result of which was used in the third equation of (\ref{QhatLhathhat})) 
and explicitly checking that Wick's theorem really holds. We will do this in
the following way. First we will assume that Wick's theorem holds for the 
normalized correlators and derive the expression for 
$\L \prod_{i=1}^M~e^{ik_iX}(z_i)\R^N_{\sigma (t_0)}$ using Wick's theorem and 
path integral arguments. Then we will show that this result is consistent with 
the one obtained in ref.\cite{dpdq} where a completely different approach was 
taken without relying on the validity of Wick's theorem.
%
%

The use of Wick's theorem immediately tells us that 
$\L \prod_{i=1}^M~e^{ik_iX}(z_i)\R^N_{\sigma (t_0)}$ should have the following
form in terms of ${\cal G}^N$.
\bea
&& \L \prod_{i=1}^M~e^{ik_iX}(z_i)\R^N_{\sigma (t_0)} \cr
\cr
&&= {\L \prod_{i=1}^M~e^{ik_iX}(z_i)~ \sigma^+ (t_0) ~\sigma^- (-t_0) \R \over
\L \sigma^+ (t_0) ~\sigma^- (-t_0) \R } \cr
\cr
&&= \eta(x_0, \vec k)~\exp\lt(-{1\over 2} \sum_{i=1}^M k_i^2 
{\cal{G}}^N_R(z_i,z_i,t_0) -\sum_{1\leq i<j\leq M} k_ik_j 
{\cal{G}}^N(z_i,z_j) \rt),    \label{e1}
\eea
where $\vec{k}$ is an $M$-dimensional vector with $k_i$'s as components and 
$x_0$ is the D.b.c of $X$ on $|t| \geq t_0$. $\eta (x_0, \vec k)$ is a 
prefactor which is not fixed by Wick's theorem.
We will determine $\eta (x_0, \vec k)$ by using path integral arguments. 
The second term in the exponential is contributed by the contractions 
between $X(z_i)$'s for different $i$'s and the first term comes from the 
self-contractions. ${\cal G}^N_R$ is the renormalized self-contraction which 
has to be defined with a specific regularization procedure. In the following 
we will first discuss this
regularization procedure to define ${\cal{G}}^N_R(z_i,z_i,t_0)$ and then 
discuss the path integral argument to determine $\eta (x_0, \vec k)$.

Let us go back to the normalized two point function ${\cal{G}}^N(z,w,t_0)$
in eq.(\ref{calG1}). Expanding the expression in powers of $(w-z)$ one can 
readily check that the short distance singularity of ${\cal{G}}^N(z,w,t_0)$
is $-2~\ln(w-z)$. This is the same as that of the free theory i.e. when the
$\sigma $ operators are absent. This is because locally the theory is always
either N.b.c. or D.b.c. (depending on where the region is chosen on the real 
line) for both of which the short distance singularity is the above one. In the
free theory one subtracts this singularity to regularize the
self-contraction. We should choose the same regularization in the presence of
$\sigma $ operators also.
This choice gives the following renormalized self-contracted two point 
function:
\bea
{\cal{G}}^N_R(z,z,t_0) &\equiv& \lim_{\delta \rightarrow 0} \lt[
{\cal{G}}^N(z,z+\delta,t_0) ~+~ 2\ln \delta \rt] \cr
\cr
&=& -2~\ln\lt[ {t_0 \over 2(t_0^2-z^2)}\rt] 
\label{calG2}
\eea

Now we will determine the prefactor $\eta (x_0, \vec k)$. In path integral 
language we replace the $\sigma $ 
operators by constraining the field $X$ to obey the proper boundary conditions
 on the real axis while integrating over the paths. With 
$z=\exp(\tau + i\sigma )$ we have the following restriction:
\bea
\begin{array}{llll}
X(\tau, \sigma )|_{\sigma =0,\pi } = x_0, & & & e^{\tau }>t_0, \\
&&& \\
\partial_{\sigma} X(\tau, \sigma )|_{\sigma =0,\pi } = 0, & & & e^{\tau }<t_0.
\end{array} \label{bc}
\eea
The basic object that one defines for computing correlation functions in path 
integral method is the following generating functional,
\bea
Z[J] = \left \L \exp \lt( i\int d\tau d\sigma J(\tau, \sigma) X(\tau, \sigma) 
\rt) \right \R.
\eea
The standard way of computing this \cite{P} is by expanding $X$ by the 
complete set of orthonormal eigenfunctions of the Laplacian operator $\nabla $,
\bea
X(\tau, \sigma ) = \sum_I \tilde x_I~\chi_I(\tau, \sigma).
\eea
One performs integrations over $\tilde x_I $'s while
computing the path integral.
Note that $\chi_0$ corresponding to the zero eigenvalue is a constant but 
$\chi_I(\tau, \sigma)$, for $I\neq 0$ are $\tau $ dependent. Hence from 
eq.(\ref{bc}) we can write,
\bea
\chi_I(\tau, \sigma )|_{\sigma =0,\pi} = 0, ~~~~~~~~~~~~~~e^{\tau}>t_0,~I\neq 0
.
\eea
Therefore for the Dirichlet regions we have from eq.(\ref{bc}),
\bea
X(\tau, \sigma )|_{\sigma =0,\pi}= \sum_I \chi_I(\tau, \sigma)|_{\sigma =0,\pi}
=\tilde x_0 \chi_0 = x_0,
\eea
which means that $\tilde x_0$ is non-dynamical and is restricted to be 
$x_0/\chi_0$. This restriction can be taken into account by inserting 
$\delta (\tilde x_0 - x_0/\chi_0) $ into the path integral. With this the 
$\tilde x_0$ integral in $Z[J]$ becomes,
\bea
Z[J] &\rightarrow& \int ~d\tilde x_0~\delta(\tilde x_0 - x_0/\chi_0)~
\exp \lt(i\tilde x_0 \chi_0 \int d\tau d\sigma J(\tau ,\sigma )\rt)\cr
\cr
&&= \exp \lt(i x_0 \int d\tau d\sigma J(\tau ,\sigma )\rt). 
\eea
To compute the correlation function in eq.(\ref{e1}) one substitutes 
$J(\tau, \sigma) = \sum_{i=1}^M k_i~\delta (\tau -\tau_i) ~\delta (\sigma 
-\sigma_i )$ which immediately suggests the following\footnote{Any other part 
of the prefactor cancels off in a normalized correlation function.}:
\bea
\eta (x_0, \vec k) = \exp (ix_0\sum_{i=1}^M k_i).
\label{eta}
\eea
Therefore using eqs. (\ref{e1}), (\ref{calG2}) and (\ref{eta}) one gets for 
$x_0=0$,
\bea
\L e^{ipX}(0) \R_{\sigma (t_0)} = (2t_0)^{-p^2-1/8},
\label{e2}
\eea
which was used in the third equation of (\ref{QhatLhathhat}).

Now we will show that the result obtained above for 
$\L \prod_{i=1}^M~e^{ik_iX}(z_i)\R^N_{\sigma (t_0)}$ (eqs. 
(\ref{e1}), (\ref{calG2}), (\ref{eta})) is consistent with the one obtained in
ref.\cite{dpdq}. In this work 
the computation was made with D.b.c. on ${\bf R}_{>0}$ and N.b.c. on 
${\bf R}_{<0}$ i.e. $\sigma^{\pm}$ were placed at $t=0$ and $t=-\infty$ 
respectively. Let us call this $\tilde z$ plane. We will be interested in the 
result on $z$ plane where $\sigma^{\pm}$ are placed at $\pm t_0$. These two
planes can be connected by an {\bf SL}(2,{\bf R}) transformation which we 
choose to be the following,
\bea
\tilde z (z) = t_0~{z-t_0 \over z+t_0}.   \label{sl2r} 
\eea
Let us then quote the result from ref.\cite{dpdq}:
\bea
\L \sigma |~\prod_{i=1}^M~ e^{ik_iX}(\tilde z_i)~|\sigma \R =
\kappa (x_0,\vec{k})~\prod_{i=1}^M (\tilde z_i)^{-k_i^2}~ 
\prod_{1\leq i<j\leq M}~\lt({\sqrt{\tilde z_i}-\sqrt{\tilde z_j} \over  
\sqrt{\tilde z_i} + \sqrt{\tilde z_j}} \rt)^{2k_ik_j}, \label{e-sigma-sigma1}
\eea
where $|\sigma \R \equiv \sigma^+(0)~|0\R$,  
$\L \sigma | \equiv \lim_{u\rightarrow 0} \L 0| I\circ \sigma^-(u) = 
\lim_{u\rightarrow 0}
u^{-1/8}~\L 0|\sigma^-(-1/u)$ with $|0\R $ being the {\bf SL}(2,{\bf R}) 
invariant vacuum and $I(u)\equiv -1/u$ being the inversion map. In 
ref.\cite{dpdq}
the above expression was found as a solution of a set of first order 
differential equations (called twisted Knizhnik-Zamolodchikov equation) 
with respect to $\tilde z_i$'s. Therefore the integration constant 
$\kappa (x_0, \vec{k})$ remained undetermined. 
We will start from eq.(\ref{e-sigma-sigma1}) and 
apply the conformal transformation eq.(\ref{sl2r}) to find the expression 
for $\L \prod_{i=1}^M~e^{ik_iX}(z_i) \R^N_{\sigma (t_0)}$. We use the 
standard rule for the conformal transformation of a correlation function of 
primary operators.
\bea
&&\L \sigma |\prod_{i=1}^M~e^{ik_iX}(\tilde z_i) |\sigma \R  \cr
\cr
&&= \lim_{u,v\rightarrow 0}~u^{-1/8}~\L \sigma(-1/u)~\prod_{i=1}^M~e^{ik_iX}
(\tilde z_i)~\sigma(v) \R \cr
\cr
\cr
&&= \lim_{u,v\rightarrow 0}~u^{-1/8}~\lt(\tilde z^{\prime}(-1/u)~\tilde 
z^{\prime}(v)\rt)^{-1/16}~\prod_{i=1}^M\lt(\tilde z^{\prime}(\tilde z_i)
\rt)^{-k_i^2} \cr
&&\quad ~ \L \sigma(z(-1/u))~\prod_{i=1}^M~e^{ik_iX}(z_i)~\sigma(z(v)) \R ,
\label{e-sigma-sigma2}
\eea
where $\displaystyle{\tilde z^{\prime}(\tilde z) \equiv {d\tilde z \over dz}}$
 expressed in terms of $\tilde z$ i.e. $\displaystyle{ \tilde z^{\prime}
(\tilde z) = {(\tilde z - t_0)^2 \over 2t_0^2 }}$. $z_i\equiv z(\tilde z_i)$.
$z(\tilde z)$ is the inverse
 function of $\tilde z(z)$ (eq.(\ref{sl2r})), namely  
$\displaystyle{z(\tilde z)=-t_0~{\tilde z + t_0 \over \tilde z - t_0}}$. 
Using 
these and eq.(\ref{e-sigma-sigma1}) and taking the limit $u,v \rightarrow 0$
we get from eq.(\ref{e-sigma-sigma2}),
\bea
\L \prod_{i=1}^M~e^{ik_iX}(z_i)\R_{\sigma (t_0)} =
\kappa(x_0,\vec{k})~(2t_0)^{-1/8}
\prod_{i=1}^M\lt({\tilde z^{\prime}_i \over \tilde z_i}\rt)^{k_i^2} ~
\prod_{1\leq i<j\leq M}~ \lt({\sqrt{\tilde z_i}-\sqrt{\tilde z_j} \over  
\sqrt{\tilde z_i} + \sqrt{\tilde z_j}} \rt)^{2k_ik_j}.  \label{e3}
\eea
Therefore the corresponding normalized correlation function will be,
\bea
\L \prod_{i=1}^M~e^{ik_iX}(z_i)\R^N_{\sigma (t_0)} =
\kappa(x_0,\vec{k})~
\prod_{i=1}^M\lt({\tilde z^{\prime}_i \over \tilde z_i}\rt)^{k_i^2} ~
\prod_{1\leq i<j\leq M}~ \lt({\sqrt{\tilde z_i}-\sqrt{\tilde z_j} \over  
\sqrt{\tilde z_i} + \sqrt{\tilde z_j}} \rt)^{2k_ik_j}.  \label{e4}  
\eea
Now using eqs.(\ref{e4}, \ref{calG1}, \ref{calG2}, \ref{sl2r}) it is 
straightforward to show that,
\bea
&& \L \prod_{i=1}^M~e^{ik_iX}(z_i)\R^N_{\sigma (t_0)} \cr
\cr
&&= 4^{(\vec{k})^2}~
\kappa(x_0,\vec{k})~\exp\lt(-{1\over 2} \sum_{i=1}^M k_i^2 
{\cal{G}}^N_R(z_i,z_i,t_0) -\sum_{1\leq i<j\leq M} k_ik_j 
{\cal{G}}^N(z_i,z_j) \rt),    \label{e5}
\eea
which is the same result as eq.(\ref{e1}) with,
\bea
\kappa (x_0, \vec{k}) = 4^{-(\vec k)^2}~ \exp (ix_0\sum_{i=1}^M k_i).
\eea

\medskip

\noindent{\bf Acknowledgement}: It is a great pleasure to thank Ashoke Sen 
for many useful discussions and guiding suggestions at various stages of this 
work. Special thanks to Justin David for many stimulating discussions and 
encouragements. I would also like to thank Ian Ellwood for discussions on 
numerical analysis. My apologies to every author whose work was not cited in 
the previous version of this paper. I would like to acknowledge the hospitality
of the Institute for Theoretical Physics, Santa Barbara where part of the work
was done.

This research was supported in part by the National 
Science Foundation under Grant No. PHY99-07949.

\end{document}